\documentclass[11pt,preprint,showpacs,preprintnumbers,amsmath,amssymb]{revtex4}

\usepackage{setspace}
\usepackage{graphicx}
\usepackage{dcolumn}
\usepackage{bm}
\usepackage{multirow}
\usepackage{graphicx}
\usepackage{amssymb}
\usepackage{float}
\usepackage{CJK}
\usepackage{caption}
\usepackage{rotating}
\usepackage{subfigure}
\usepackage{relsize}
\usepackage{diagbox}
\usepackage{epstopdf}

\begin{document}

\title{Studying the $D_1D$ molecule in the Bethe-Salpeter equation approach}

\author{Zhen-Yang Wang \footnote{e-mail: wangzhenyang@nbu.edu.cn}}
\affiliation{\scriptsize{Physics Department, Ningbo University, Zhejiang 315211, China}}

\author{Jing-Juan Qi \footnote{Corresponding author, e-mail: qijj@mail.bnu.edu.cn}}
\affiliation{\scriptsize{Junior College, Zhejiang Wanli University, Zhejiang 315101, China}}

\author{Jing Xu \footnote{e-mail: xj2012@mail.bnu.edu.cn}}
\affiliation{\scriptsize{Department of Physics, Yantai University, Yantai 264005, China}}

\author{Xin-Heng Guo \footnote{Corresponding author, e-mail: xhguo@bnu.edu.cn}}
\affiliation{\scriptsize{College of Nuclear Science and Technology, Beijing Normal University, Beijing 100875, China}}

\date{\today}

\begin{abstract}
We study the possible bound states of the $D_1D$ system in the Bethe-Salpeter (BS) formalism in the ladder and instantaneous approximations. By solving the BS equation numerically with the kernel containing one-particle exchange diagrams and introducing three different form factors (monopole, dipole, and exponential form factors) at the vertices, we investigate whether the isoscalar and isovector $D_1D$ bound states may exist, respectively. We find that $Y(4260)$ could be accommodated as a $D_1D$ molecule, whereas the interpretation of $Z_2^+(4250)$ as a $D_1D$ molecule is disfavored. The bottom analog of $Y(4260)$ may exist but that of $Z_2^+(4250)$ does not.
\end{abstract}


\maketitle
\section{Introduction}
\label{intro}

The charmonium-like state $Y(4260)$ [or named as $\psi(4260)$] was first observed by $BABAR$ Collaboration in the initial-state radiation process $e^+e^-\rightarrow \gamma_{ISR}J/\psi\pi^+\pi^-$  in 2005 \cite{Aubert:2005rm}, and then immediately confirmed by CLEO \cite{He:2006kg} and Belle \cite{Yuan:2007sj} Collaborations in the same process. While BES$\mathrm{\uppercase\expandafter{\romannumeral3}}$ Collaboration observed $Y(4260)$ afterwards, it also reported a stunning particle-$Z_c(3900)$ in $e^+e^-\rightarrow Y(4260)\rightarrow J/\psi \pi^+\pi^-$ process \cite{Ablikim:2013mio}. The average mass and width of $Y(4260)$ are $M = 4230\pm8$ MeV and $\Gamma=55\pm19$ MeV in PDG \cite{Tanabashi:2018oca}, respectively. In 2007, BES$\mathrm{\uppercase\expandafter{\romannumeral3}}$ Collaboration performed a precise cross section measurement of $e^+e^-\rightarrow J/\psi \pi^+\pi^-$ for c.m. energies from $\sqrt{s}$ = 3.77 to 4.60 GeV and observed two resonant structures, one with a mass of (4222.0$\pm$3.1$\pm$1.4) MeV and a width of (44.1 $\pm$4.3$\pm$2.0) MeV and the other with a mass of (4320.0$\pm$ 10.4$\pm$7.0) MeV and a width of ($101.4^{+25.3}_{-19.7}\pm$10.2) MeV \cite{Ablikim:2016qzw}. The first resonance (named as $Y(4220)$) agrees with the $Y(4260)$ resonance reported by previous experiments. Recently, the first experimental evidence for open-charm production ($e^+e^-\rightarrow Y(4220)\rightarrow\pi^+D^0D^{\ast -}$) associated with the $Y(4220)$ state was observed by BES$\mathrm{\uppercase\expandafter{\romannumeral3}}$ \cite{Ablikim:2018vxx}.

With $Y(4260)$ being produced via $e^+e^-$ annihilation, its $J^{PC}$ should be $1^{--}$. Since $Y(4260)$ is well above the $D\bar{D}$ threshold, it should have a large phase space to decay into charmed meson pairs. However, unlike the charmonium states $\psi(4040)$, $\psi(4160)$, and $\psi(4415)$ in the same mass range which decay predominantly into open charm final states, $Y(4260)$ only show a strong coupling to $\pi^+\pi^-J/\psi$, and has not been observed in any open charm decay channels like $D\bar{D}$, $D^\ast\bar{D}+c.c.$, and $D^\ast\bar{D}^\ast$ \cite{Abe:2006fj,Pakhlova:2008zza,Aubert:2009aq,CroninHennessy:2008yi}. Furthermore, $Y(4260)$ does not fit in the conventional charmonium spectroscopy \cite{Brambilla:2010cs}. These features suggest a complicated substructure of $Y(4260)$. In fact, $Y(4260)$ has stimulated lots of studies with different theoretical structure assumptions, including hybrid state \cite{Zhu:2005hp,Kou:2005gt}, tetraquark state \cite{Ebert:2008kb,Ali:2011qi,Dias:2012ek}, charmonium and tetraquark mixing state \cite{Albuquerque:2015nwa,Wang:2016mmg}, hadronic molecule of $\chi_{c1}\rho^0$ \cite{Liu:2005ay}, $\chi_{c1}\omega$ \cite{Yuan:2005dr}, $D_0\bar{D}^\ast$ \cite{Albuquerque:2008up}, $J/\psi f_0(980)$ \cite{Albuquerque:2011ix}, $D_1D$ \cite{Ding:2008gr,Liu:2013vfa,Cleven:2013mka,Qin:2016spb,Cleven:2016qbn,Xue:2017xpu,Chen:2019mgp} or $J/\psi K\bar{K}$ \cite{MartinezTorres:2009xb}, and baryonium state \cite{Qiao:2005av}. Since the mass of $Y(4260)$ is only about 29 MeV below the threshold of $D_1 D$  which is the first open-charm $S$-wave channel coupling a state with $J^{PC}= 1^{--}$, the $D_1 D$ molecule is a good candidate for the structure of $Y(4260)$, which has remained controversial until now.

Years ago Belle Collaboration observed two charged resonance-like structures, $Z_1^+(4051)$ and $Z_2^+(4250)$, with the significance of more than 5$\sigma$ in the $\chi_{c1}\pi^+$ mass distribution in $B\rightarrow K^-\pi^+\chi_{c1}$ decays \cite{Mizuk:2008me}. Their Breit-Wigner masses and widths are $M_1=4051\pm14^{+20}_{-41}$ MeV, $\Gamma_1=82^{+21+47}_{-39-61}$ MeV, and $M_2=4248^{+44+180}_{-29-35}$ MeV, $\Gamma_2=177^{+54+316}_{-39-61}$ MeV, respectively. As both $Z_1^+(4051)$ and $Z_2^+(4250)$ carry one unit electric charge, if these states do exist they cannot be traditional $q\bar{q}$ quark bound states. In 2012, $BABAR$ Collaboration searched for these resonances in the $\bar{B}^0\rightarrow \chi_{c1}K^-\pi^+$ and $B^+\rightarrow \chi_{c1}K_S^0\pi^+$ decays and did not find any evidence of them \cite{Lees:2011ik}. In 2013, with more than twice the Belle and $BABAR$ cumulative events and using the same analysis strategy as that of $BABAR$, LHCb Collaboration did not support the evidence for the existence of these two resonances in the $B^0\rightarrow \chi_{c1}K^+\pi^-$ channel \cite{Sbordone:2013exb}. Since $\pi^+$ is an isovector meson with negative $G$ parity and $\chi_{c1}$ is a isospin singlet with positive $G$ parity, if $Z_1^+(4051)$ and $Z_2^+(4250)$ exist, their quantum numbers should be $I^G=1^-$. In theoretical studies, $Z_2^+(4250)$ has been interpreted as a $D_1\bar{D}$ molecular state \cite{Lee:2008tz} or a tetraquark state \cite{Wang:2008af,Deng:2015lca}.

$Y(4260)$ and $Z_2^+(4250)$ provide a great opportunity for understanding the strong interaction dynamics inside a hadron with moleculer inner structure assumptions since their masses are close to the $D_1D$ threshold. We will systematically study the $D_1D$ molecular state in the Bethe-Salpeter (BS) equation approach with three different form factors at the interaction verties, We will investigate the $S$-wave $D_1D$ systems with isospins $I$ =0 and 1 being both considered. We will vary the binding energy $E_b=M-M_{D_1}-M_D$ (where $M$ is mass of the bound state) in a wide range and search for all the possible solutions with the cutoff parameter $\Lambda$ in the form factor in a reasonable interval. Through this process, we will naturally check whether $Y(4260)$ and $Z_2^+(4250)$ may exist as a $S$-wave $D_1D$ molecular state. The possible $B_1B$ molecular state will also been studied in our work.

In the rest of the manuscript we will proceed as follows. In Sec. \ref{sect-BS-PV}, we will establish the BS equation for the bound state of an axial-vector meson ($D_1$ or $B_1$) and a pseudoscalar meson ($D$ or $B$). Then the numerical results for the $D_1D$ and $B_1B$ systems will be presented in Sec. \ref{NR}. In Sec. \ref{su} we will present a summary of our results.

\section{The BS formalism for $D_1D$ system}
\label{sect-BS-PV}
As discussed in Ref. \cite{Ding:2008gr}, the flavor wave functions of $Y(4250)$ and $Z_2^+(4250)$ are
\begin{equation}
|Y(4260)\rangle=\frac12[|D_1^0\bar{D}^0\rangle+|D_1^+D^-\rangle|-|D^0\bar{D}_1^0\rangle-|D^+D_1^-\rangle],
\end{equation}
and
\begin{equation}
|Z_2^+(4250)\rangle=\frac{1}{\sqrt{2}}[|D_1^+\bar{D}^0\rangle+|D^+\bar{D}_1^0\rangle],
\end{equation}
respectively.

Based on the picture that $Y(4250)$ and $Z_2^+(4250)$ are composed of an axial-vector meson ($D_1$) and a meson ($D$), its BS wave function is defined as
\begin{equation}
  \chi^\mu\left(x_1,x_2,P\right) = \langle0|TD_1^\mu(x_1)D(x_2)|P\rangle,
\end{equation}
where $D_1(x_1)$ and $D(x_2)$ are the field operators of the axial-vector meson $D_1$ and pseudoscalar meson $D$ at space coordinates $x_1$ and $x_2$, respectively, $P=Mv$ is the total momentum of $Y(4250)$ or $Z_2^+(4250)$ and $v$ is its velocity. Let $m_{D_1}$ and $m_D$ be the masses of the $D_1$ and $D$ mesons, respectively, $p$ be the relative momentum of the two constituents, and define $\lambda_1=m_{D_1}/(m_{D_1}+m_D)$, $\lambda_2=m_D/(m_{D_1}+m_D)$. The BS wave function in momentum space is defined as
\begin{equation}\label{PV-momentum-BS-function}
 \chi^\mu_P(x_1,x_2,P) = e^{-iPX}\int\frac{d^4p}{(2\pi)^4}e^{-ipx}\chi^\mu_P(p),
\end{equation}
where $ X = \lambda_1x_1 + \lambda_2x_2$ is the coordinate of the center of mass and $x = x_1 - x_2$. The momentum of the $D_1$ meson is $p_1=\lambda_1P+p$ and that of the $D$ meson is $p_2=\lambda_2P-p$.

It can be shown that the BS wave function of the $D_1D$ system satisfies the following BS equation \cite{Lurie}:
\begin{equation}\label{PV-BS-equation}
  \chi^\mu_{P}(p)=S^{\mu\nu}_{D_1}(p_1)\int\frac{d^4q}{(2\pi)^4}K_{\nu\lambda}(P,p,q)\chi^\lambda_{P}(q)S_{D}(p_2),
\end{equation}
where $S^{\mu\nu}_{D_1}(p_1)$ and $S_{D}(p_2)$ are the propagators of $D_1$ and $D$, respectively, and $K_{\nu\lambda}(P,p,q)$ is the kernel, which is defined as the sum of all the two particle irreducible diagrams with respect to $D_1$ and $D$ mesons. For convenience, in the following we use the variables $p_l (=p\cdot v)$ and $p_t(=p- p_lv)$ as the longitudinal and transverse projections of the relative momentum ($p$) along the bound state momentum ($P$), respectively. Then the propagator of the $D_1$ meson can be expressed as
\begin{equation}\label{D1-propagator}
  S^{\mu\nu}_{D_1}(\lambda_1P+p)=\frac{-i\left(g^{\mu\nu}-p_1^\mu p_1^\nu/m^2_{D_1}\right)}{\left(\lambda_1M+p_l\right)^2-\omega_1^2+i\epsilon},
\end{equation}
and the propagator of the pseudoscalar $D$ meson has the form
\begin{equation}\label{D-propagator}
  S_D(\lambda_2P-p)=\frac{i}{\left(\lambda_2M-p_l\right)^2-\omega_2^2+i\epsilon},
\end{equation}
where $\omega_{1(2)} = \sqrt{m_{1(2)}^2+p_t^2}$ (we have defined $p_t^2=-p_t\cdot p_t$). The momentum of $p_1$ in the numerator of Eq. (\ref{D1-propagator}) are represented by $p_l$ and $p_t$ in the following:
\begin{equation}
 p_1=(\lambda_1 M+p_l)v+p_t.
\end{equation}

In the BS equation approach, the interaction between $D_1$ and $D$ mesons can be due to the light vector-meson ($\rho$ and $\omega$) and the light scalar-meson ($\sigma$) exchanges. Based on the heavy quark symmetry and the chiral symmetry, the relevant effective Lagrangian used in this work is shown in the following \cite{Ding:2008gr}:
\begin{equation}
\begin{split}\label{D1D-Lagrangian}
  \mathcal{L}_{DD\sigma} =& g_{DD\sigma}D_aD_a^\dag\sigma+g_{\bar{D}\bar{D}\sigma}\bar{D}_a\bar{D}_a^\dag\sigma,\\
  \mathcal{L}_{D_1D_1\sigma} =& g_{D_1D_1\sigma}D_{1a}^\mu D_{1a\mu}^\dag\sigma+g_{\bar{D}_1\bar{D}_1\sigma}\bar{D}_{1a}^\mu\bar{D}_{1a\mu}^\dag\sigma,\\
  \mathcal{L}_{DD_1\sigma} =& g_{DD_1\sigma}D_{1a}^\mu D_{a}^\dag\partial_\mu\sigma+g_{\bar{D}\bar{D}_1\sigma}\bar{D}_{1a}^\mu\bar{D}_{a}^\dag\partial_\mu\sigma+H.c.,\\
  \mathcal{L}_{DDV} =&ig_{DDV}(D_b\overleftrightarrow{\partial}_\mu D_a^\dag)V_{ba}^\mu+ig_{\bar{D}\bar{D}V}(\bar{D}_b\overleftrightarrow{\partial}_\mu\bar{D}_a^\dag)V_{ba}^\mu,\\
  \mathcal{L}_{D_1D_1V} =&ig_{D_1D_1V}(D^\nu_{1b}\overleftrightarrow{\partial}_\mu D_{1a\nu}^\dag)V_{ba}^\mu+ig'_{D_1D_1V}(D_{1b}^\mu D_{1a}^{\nu\dag}-D_{1a}^{\mu\dag}D_{1a}^\nu)(\partial_\mu V_\nu-\partial_\nu V_\mu)_{ba}\\
  &+ig_{\bar{D}_1\bar{D}_1V}(\bar{D}_{1b\nu}\overleftrightarrow{\partial}_\mu \bar{D}_{1a}^{\nu\dag})V_{ab}^\mu+ig'_{\bar{D}_1\bar{D}_1V}(\bar{D}_{1b}^\mu \bar{D}_{1a}^{\nu\dag}-\bar{D}_{1a}^{\mu\dag}\bar{D}_{1b}^\nu)(\partial_\mu V_\nu-\partial_\nu V_\mu)_{ab}\\
  \mathcal{L}_{DD_1V}=&g_{DD_1V}D_{1b}^\mu V_{\mu ba}D_a^\dag+g'_{DD_1V}(D_{1b}^\mu\overleftrightarrow{\partial}^\nu D_a^\dag)(\partial_\mu V_\nu-\partial_\nu V_\mu)_{ba}\\
  &+g_{\bar{D}\bar{D}_1V}\bar{D}_a^\dag V_{\mu ba}\bar{D}_{1b}^\mu+g'_{\bar{D}\bar{D}_1V}(\bar{D}_{1b}^\mu\overleftrightarrow{\partial}^\nu \bar{D}_a^\dag)(\partial_\mu V_\nu-\partial_\nu V_\mu)_{ba}+H.c.,\\
  \end{split}
\end{equation}
where $a$ and $b$ are represent the light flavor quark, $V_\mu$ is a $3\times3$ Hermitian matrix containing $\rho$, $\omega$, $K^\ast$, and $\phi$:
\begin{eqnarray}
V&=&\left(\begin{array}{ccc}
\frac{\rho^{0}}{\sqrt{2}}+\frac{\omega}{\sqrt{2}}&\rho^{+}&K^{*+}\\
\rho^{-}&-\frac{\rho^{0}}{\sqrt{2}}+\frac{\omega}{\sqrt{2}}&
K^{*0}\\
K^{*-} &\bar{K}^{*0}&\phi
\end{array}\right).\label{vector}
\end{eqnarray}
The coupling constants involved in Eq. (\ref{D1D-Lagrangian}) are related to each other as follows \cite{Ding:2008gr}:
\begin{equation}
\begin{split}
g_{DD\sigma}&=g_{\bar{D}\bar{D}\sigma}=-2g_\sigma m_D, \quad\quad g_{D_1D_1\sigma}=g_{\bar{D}_1\bar{D}_1\sigma}=-2g^{''}_\sigma m_{D_1},\\
g_{DD_1\sigma}&=g_{\bar{D}\bar{D}_1\sigma}=-\frac{2\sqrt{6}}{3}\frac{h'_\sigma}{f_\pi}\sqrt{m_Dm_{D_1}},\\
g_{DDV}&=-g_{\bar{D}\bar{D}V}=\frac{1}{\sqrt{2}}\beta g_V,\quad\quad g_{D_1D_1V}=-g_{\bar{D}_1\bar{D}_1V}=\frac{1}{\sqrt{2}}\beta_2 g_V,\\
g'_{D_1D_1V}&=-g'_{\bar{D}_1\bar{D}_1V}=\frac{5\lambda_2g_V}{3\sqrt{2}}m_{D_1},\\
g_{DD_1V}&=-g_{\bar{D}\bar{D}_1V}=-\frac{2}{\sqrt{3}}\zeta_1 g_V\sqrt{m_Dm_{D_1}},\quad\quad g'_{DD_1V}=-g'_{\bar{D}\bar{D}_1V}=\frac{1}{\sqrt{3}}\mu_1 g_V,\\
\end{split}
\end{equation}
with
\begin{equation}
\begin{split}
g_\sigma&=-\frac{g_\pi}{2\sqrt{6}},\quad\quad h_\sigma=\frac{g_A}{\sqrt{3}},\\
\end{split}
\end{equation}
where $f_\pi=132$ MeV, $g_\pi=3.73$ and $g_A=0.6$ \cite{Bardeen:2003kt}.  As in Ref. \cite{Liu:2008xz}, we take $|g^{''}_\sigma|=|g_\sigma|$ and $|h'_\sigma|=|h_\sigma|$ approximately when performing the numerical analysis. The parameters $\beta_2g_V$ and $\lambda_2g_V$ are given by $2g_{\rho NN}$ and $\frac{3}{10m_N}(g_{\rho NN}+f_{\rho NN})$, respectively, where $g_{\rho NN}^2/4\pi=0.84$ and $f_{\rho NN}/g_{\rho NN}=6.10$ \cite{Wang:2019aoc}. As to the two parameters $\zeta_1$ and $\mu_1$ involved in the coupling constants, the information about them is very scarce and these two parameters have not been determined. However in the heavy quark limit, we can roughly assume that the coupling constants $g_{DD_1V}$ and $g'_{DD_1V}$ are equal to $g_{D^\ast D_0V}$ (=$\zeta g_V\sqrt{2m_{D^\ast}m_{D^0}}$) and $g'_{D^\ast D_0V}$ (=$1/\sqrt{2}\mu g_V$), respectively. The parameters $\mu=0.1$ $\mathrm{GeV}^{-1}$ and $\zeta=0.1$ are taken in Ref. \cite{Casalbuoni:1996pg}. In our calculations, we will vary $\mu_1$ from 0.05 to 0.5 $\mathrm{GeV}^{-1}$ and $\zeta_1$ from 0.05 to 0.5, while searching for possible solutions of the $D_1D$ bound states.

In the following we will given the kernel for the BS equation in the ladder approximation. There have been some studies on the legitimacy of applying the ladder approximation in the BS equation \cite{Gross:1982nz,Theussl:1999xq,Guo:2007mm}. In Ref. \cite{Gross:1982nz} it was shown that including only ladder graphs in the scalar-scalar system cannot lead to the correct one-body limit, and to solve these problems, at least crossed ladder graphs should be included. In addition, from the naive perspective, for a large coupling constant, the ladder approximation is not legitimate \cite{Theussl:1999xq}. However, there is a significant difference between our work and that studied in Refs. \cite{Theussl:1999xq}, in which the mass of the exchanged particle ($\mu$) is very small compared to the mass of the constituent particle ($m$) with $\mu/m$ = 0.15. The exchanged particles in our work are $\sigma$, $\rho$ and $\omega$. In Ref. \cite{Guo:2007mm}, the authors studied the $K\bar{K}$ bound states in the BS equation, they found that when $\rho$ is exchanged the ratio of the contribution from the crossed graph to that from the ladder one is less than 15$\%$ and the result is almost the same when $\omega$ is exchanged. Therefore, the ladder approximation is a good one which should not affect our qualitative conclusions.

\begin{figure}[htbp]
\centering
\subfigure[]{
\begin{minipage}[t]{0.3\linewidth}
\centering
\includegraphics[width=1.5in]{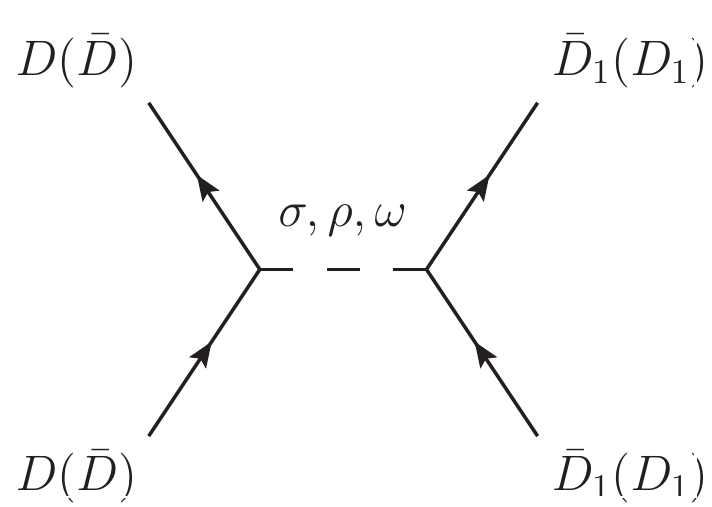}
\end{minipage}%
}%
\subfigure[]{
\begin{minipage}[t]{0.3\linewidth}
\centering
\includegraphics[width=1.6in]{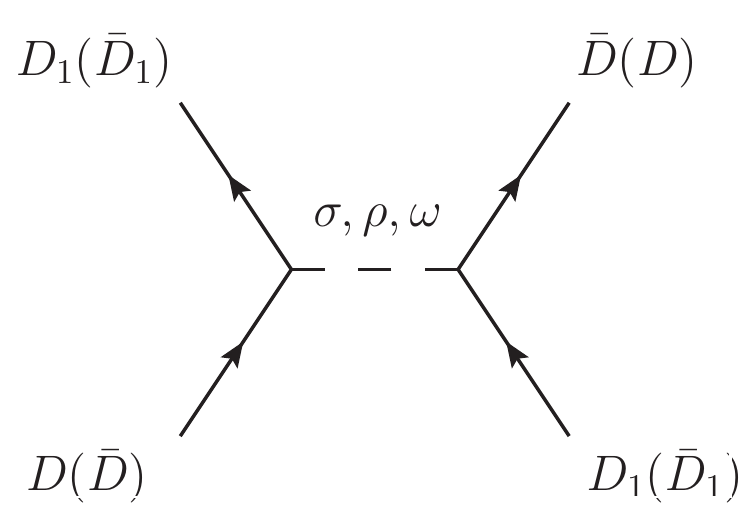}
\end{minipage}
}%
\centering
\caption{The direct-channel (a) and cross-channel (b) Feynman diagrams for the $D_1D$ system at the tree level.}
\label{feynman}
\end{figure}

Then, at the tree level, in the $t$-channel the kernel for the BS equation of $D_1D$ in the lader approximation includes the following terms (see Figs. \ref{feynman}(a) and \ref{feynman}(b) for direct and crossed channels, respectively):
\begin{equation}\label{D1D-kernel}
\begin{split}
  K^{\tau\sigma}_{direct}(P,p,q;m_\sigma)=&-(2\pi)^4\delta^4(p'_1+p'_2-p_1-p_2)c_I g_{D_1D_1\sigma}g_{DD\sigma}\Delta(k,m_\sigma)g^{\tau\sigma},\\
  K^{\tau\sigma}_{direct}(P,p,q;m_V)=&-(2\pi)^4\delta^4(p'_1+p'_2-p_1-p_2)c_I \Big\{g_{D_1D_1V}g_{DDV}(p_1+q_1)_\gamma(p_2+q_2)_\rho g^{\tau\sigma}\\
  &\times\Delta^{\rho\gamma}(k,m_V)+g'_{D_1D_1V}g_{DDV}(p_2+q_2)_\rho\left[k^\tau\Delta^{\rho\sigma}(k,m_V)-k^\sigma\Delta^{\rho\tau}(k,m_V)\right]\Big\},\\
  K^{\tau\sigma}_{crossed}(P,p,q;m_\sigma)=&-(2\pi)^4\delta^4(p'_1+p'_2-p_1-p_2)c_Ig_{DD_1\sigma}^2k^\tau k^\sigma \Delta(k,m_\sigma),\\
  K^{\tau\sigma}_{crossed}(P,p,q;m_V)=&-(2\pi)^4\delta^4(p'_1+p'_2-p_1-p_2)\Big\{g_{DD_1V}^2\Delta^{\tau\sigma}(k,m_V)\\
  &+g_{DD_1V}g'_{DD_1V}(q_1+p_2)^\gamma\left[k_\gamma\Delta^{\tau\sigma}(k,m_V)-k^\tau\Delta_\gamma^\sigma(k,m_V)\right]\\
  &+g_{DD_1V}g'_{DD_1V}(p_1+q_2)^\rho\left[k_\rho\Delta^{\tau\sigma}(k,m_V)-k^\tau\Delta_\rho^\sigma(k,m_V)\right]\\
  &+g_{DD_1V}^{'2}(p_1+q_2)^\rho(q_1+p_2)^\gamma\big[k_\rho k_\gamma\Delta^{\tau\sigma}(k,m_V)-k_\rho k^\sigma\Delta^\tau_\gamma(k,m_V)\\
  &-k^\tau k_\gamma\Delta_\rho^\sigma(k,m_V)+k^\tau k^\sigma\Delta_{\rho\gamma}(k,m_V)\big]\Big\},
\end{split}
\end{equation}
where $m_V$ represent the masses of the exchanged light vector mesons $\rho$ and $\omega$ , $c_I$ is the isospin coefficient: $c_0 =3, 1, 1$ and $c_1 =-1, 1, 1$ for $\rho$, $\omega$, and $\sigma$, respectively, $\Delta$ and $\Delta^{\mu\nu}$ represent the propagators for the scalar and vector mesons, respectively.

In order to describe the phenomena in the real world, we should include a form factor at each interacting vertex of hadrons to include the finite-size effects of these hadrons. For the meson-exchange case, the form factor is assumed to take the following form \cite{Chen:2017vai}:
\begin{equation}\label{form-factor}
\begin{split}
  F_M(k)&=\frac{\Lambda_M^2-m^2}{\Lambda_M^2-k^2}, \\
  F_D(k)&=\frac{(\Lambda_D^2-m^2)^2}{(\Lambda_D^2-k^2)^2}, \\
  F_E(k)&=e^{(k^2-m^2)/\Lambda_E^2}, \\
\end{split}
\end{equation}
where $\Lambda$, $m$ and $k$ represent the cutoff parameter, the mass of the exchanged meson and the momentum of the exchanged meson, respectively. The value of $\Lambda$ is near 1 GeV which is the typical chiral symmetry breaking scale.

In general, for an axial-vector meson ($D_1$) and a pseudoscalar meson ($D$) bound state, the BS wave function $\chi_P^\mu(p)$ has the following form:
\begin{equation}
\chi_P^\mu(p)=f_0(p)p^\mu+f_1(p)P^\mu+f_2(p)\epsilon^\mu+f_3(p)\varepsilon^{\mu\nu\alpha\beta}p_\alpha P_\beta\epsilon_\nu,
\end{equation}
where $f_i(p)$ $(i = 0,1,2,3)$ are Lorentz-scalar functions and $\epsilon^\mu$ represents the polarization vector of the bound state. After considering the constraints imposed by parity and Lorentz transformations, it is easy to prove that $\chi_P^\mu(p)$ can be simplified as
\begin{equation}\label{BS-wave-function}
  \chi_P^\mu(p)=f(p)\varepsilon^{\mu\nu\alpha\beta}p_\alpha P_\beta\epsilon_\nu,
\end{equation}
where the function $f(p)$ contains all the dynamics.

In the following derivation of the BS equation, we will apply the instantaneous approximation, in which the energy exchanged between the constituent particles of the binding system is neglected. In our calculation we choose the absolute value of the binding energy $E_b$ of the $D_1D$ system (which is defined as $E_b=M-m_1-m_2$) less than 60 MeV. In this case the exchange of energy between the constituent particles can be neglected.

Substituting Eqs. (\ref{D1-propagator}), (\ref{D-propagator}), (\ref{D1D-kernel}) and (\ref{form-factor}) into Eq. (\ref{PV-BS-equation}) and using the covariant instantaneous approximation in the kernel, $p_l=q_l$, one obtains the folowing expression:
\begin{equation}\label{4-dimension-BS-equation}
  \begin{split}
   &f(p) =\frac{ic_I}{3[(\lambda_1M+p_l)^2-\omega_1^2+i\epsilon][(\lambda_2M-p_l)^2-\omega_2^2+i\epsilon]}\int\frac{d^4q}{(2\pi)^4}\\
    &\Bigg\{g_{\bar{D}_1\bar{D}_1\sigma}g_{DD\sigma}\frac{1}{-(p_t-q_t)^2-m_\sigma^2}\left[\frac{(\lambda_1M+p_l)^2-p_t^2}{m_1^2}-\frac{M(\lambda_1M+p_l)(p_l(\lambda_1M+p_l)-p_t\cdot q_t)}{m_1^2Mp_l}-3\right]\\
    &-g_{\bar{D}_1\bar{D}_1V}g_{DDV}\frac{1}{-(p_t-q_t)^2-m_V^2}\bigg[-\frac{((\lambda_1M+p_l)^2-p_t^2)(p_t^2-q_t^2)^2}{m_1^2m_V^2}+\frac{3(p_t^2-q_t^2)^2}{m_V^2}\\
    &-\frac{(p_t^2-q_t^2)^2M(\lambda_1M+p_l)(p_l(\lambda_1M+p_l)-p_t\cdot q_t)}{m_1^2m_V^2Mp_l}-3(4(\lambda_1M+p_l)(\lambda_2M-p_l)+(p_t+q_t)^2)\\
    &-\frac{M(\lambda_1M+p_l)(p_l(\lambda_1M+p_l)-p_t\cdot q_t)(4(\lambda_1M+p_l)(\lambda_2M-p_l)+(p_t+q_t)^2)}{m_1^2M p_l}\\
    &+\frac{((\lambda_1M+p_l)^2-p_t^2)(4(\lambda_1M+p_l)(\lambda_2M-p_l)+(p_t+q_t)^2)}{m_1^2}\bigg]\\
    &+g'_{\bar{D}_1\bar{D}_1V}g_{DDV}\frac{1}{-(p_t-q_t)^2-m_V^2}\bigg[\frac{2(p_t^2+p_t\cdot q_t)M(\lambda_2M-p_l)(p_l(\lambda_1M+p_l)-p_t\cdot q_t)}{m_1^2Mp_l}\\
    &-\frac{2(p_t\cdot q_t-q_t^2)M(\lambda_2M-p_l)}{Mp_l}\bigg]\\
    &-g_{\bar{D}_1\bar{D}\sigma}g_{D_1D\sigma}\frac{1}{-(p_t-q_t)^2-m_\sigma^2}\bigg[\frac{(p_t^2-p_t\cdot q_t)^2}{m_1^2}+(p_t-q_t)^2\bigg]\\
    &+g_{\bar{D}_1\bar{D}V}g_{D_1DV}\frac{1}{-(p_t-q_t)^2-m_V^2}\bigg[\frac{(p_t^2-p_t\cdot q_t)^2}{m_1^2m_V^2}+\frac{(p_t-q_t)^2}{m_V^2}-\frac{(\lambda_1M+p_l)^2-p_t^2}{m_1^2}\\
    &+\frac{M(\lambda_1M+p_l)(p_l(\lambda_1M+p_l)-p_t\cdot q_t)}{m_1^2Mp_l}+3\bigg]\\
    &-g_{\bar{D}_1\bar{D}V}g'_{D_1DV}\frac{1}{-(p_t-q_t)^2-m_V^2}\bigg[\frac{((\lambda_1M+p_l)^2-p_t^2)(p_t-q_t)^2}{m_1^2}+\frac{(p_t^2-p_t\cdot q_t)(M(\lambda_1M+p_l)+p_t^2-p_t\cdot q_t)}{m_1^2}\\
    &-\frac{(p_t-q_t)^2M(\lambda_1M+p_l)(p_l(\lambda_1M+p_l)-p_t\cdot q_t)}{m_1^2M p_l}-\frac{(p_t^2-p_t\cdot q_t)M^2(p_l(\lambda_1M+p_l)-p_t\cdot q_t)}{m_1^2Mp_l}\\
    &-2(p_t-q_t)^2+\frac{(p_t\cdot q_t-q_t^2)M^2}{Mp_l}\bigg]\\
    &+g'_{\bar{D}_1\bar{D}V}g_{D_1DV}\frac{1}{-(p_t-q_t)^2-m_V^2}\bigg[\frac{(p_t^2-p_t\cdot q_t)M^2(p_l(\lambda_1M+p_l)-p_t\cdot q_t)}{m_1^2Mp_l}+\frac{((\lambda_1M+p_l)^2-p_t^2)(p_t-q_t)^2}{m_1^2}\\
    &-\frac{(p_t^2-p_t\cdot q_t)(M(\lambda_1M+p_l)-p_t^2+p_t\cdot q_t)}{m_1^2}-\frac{(p_t-q_t)^2M(\lambda_1M+p_l)(p_l(\lambda_1M+p_l)-p_t\cdot q_t)}{m_1^2Mp_l}\\
    &-\frac{(p_t\cdot q_t-q_t^2)M^2}{Mp_l}-2(p_t-q_t)^2\bigg]\\
    &+g'_{\bar{D}_1\bar{D}V}g'_{D_1DV}\frac{1}{-(p_t-q_t)^2-m_V^2}\bigg[\frac{(M^2+(p_t-q_t)^2)(p_t^2-p_t\cdot q_t)}{m_1^2}-\frac{(p_t\cdot q_t-q_t^2)(p_t-q_t)^2M^2}{M p_l}\\
    &+\frac{(p_t^2-p_t\cdot q_t)(p_t-q_t)^2M^2(p_l(\lambda_1M+p_l)-p_t\cdot q_t)}{m_1^2Mp_l}+\frac{(p_t^2-p_t\cdot q_t)(p_t-q_t)^2(M(\lambda_1M+p_l)+p_t^2-p_t\cdot q_t)}{m_1^2}\\
    &-\frac{(p_t-q_t)^4((\lambda_1M+p_l)^2-p_t^2)}{m_1^2}-\frac{(p_t-q_t)^4(p_l(\lambda_1M+p_l)-p_t\cdot q_t)M(\lambda_1M+p_l)}{m_1^2Mp_l}\\
    &-\frac{(p_t^2-p_t\cdot q_t)(p_t-q_t)^2(M(\lambda_1M+p_l)-p_t^2+p_t\cdot q_t)}{m_1^2}-(p_t-q_t)^4-(p_t-q_t)^2(M^2+(p_t-q_t)^2)\bigg]\Bigg\}f(q).
  \end{split}
\end{equation}

\section{Numerical results}
\label{NR}
In this section, we will solve the BS equations numerically for the $D_1 D$ systems with $I=0$ and $I=1$ based on the formulas presented in Sec. \ref{sect-BS-PV} and study whether the $S$-wave $D_1 D$ molecular states exist. We first need to reduce the BS equation (\ref{4-dimension-BS-equation}) to a one-dimensional form. By choosing the appropriate contour and performing the integration over $p_l$ on both sides through applying the residue theorem, we can reduce the BS equation (\ref{4-dimension-BS-equation}) to a three-dimensional form. The corresponding BS wave function is in fact rotationally invariant, i.e. $\tilde{f}(\mathbf{p}_t)$ (where $\tilde{f}(\mathbf{p}_t)=\int dp_lf(p)$) depends only on the norm of the three momentum, $|\mathbf{p}_t|$. Therefore, after completing the azimuthal integration, we can obtain the one-dimensional BS equation.

The BS wave functions for $D_1 D$ systems with $I=0$ and $I=1$ were solved numerically in our previous work by discretizing the integration region (0,$\infty$) into $n$ pieces \cite{Wang:2019ehs} ($n$ is chosen to be sufficiently large and we use $n$-point Gauss quadrature rule to evaluate the integrals). Then the BS wave function can be written as an $n$-dimention vector. The  coupled integral BS equation becomes a matrix equation
\begin{equation}
\tilde{f}(|\mathbf{p}_t|_n)=A(|\mathbf{p}_t|_n, |\mathbf{q}_t|_n)\cdot\tilde{f}(|\mathbf{q}_t|_n),
\end{equation}
where $A(|\mathbf{p}_t|_n, |\mathbf{q}_t|_n)$ corresponding to the coefcients in Eq. (\ref{4-dimension-BS-equation}). Generally, $|\mathbf{p}_t|$ varies from 0 to +$\infty$ and $\tilde{f}(|\mathbf{p}_t|)$ will decrease to zero when $|\mathbf{p}_t|\rightarrow\infty$. To apply the Gaussian quadrature rule, we need to convert the Gaussian integration nodes into the physical values for $|\mathbf{q}_t|$, which can be done using the following equation:
\begin{equation}
|\mathbf{q}_t|=\epsilon+w\log\left[1+y\frac{1+t}{1-t}\right],
\end{equation}
where $\epsilon$ is a parameter introduced to avoid divergence in numerical calculations, $w$ and $y$ are parameters used in controlling the slopes of wave functions and finding the proper solutions for these functions, and $t$ varies from -1 to 1. One can then obtain the numerical results of $\tilde{f}(|\mathbf{p}_t|)$ by requiring the eigenvalue of the eigenvalue equation to be 1.0.

It can be seen from Eq. (\ref{4-dimension-BS-equation}) that there is only one free parameter in our model, the cutoff $\Lambda$, which can not be uniquely determined and has various forms phenomenologically. It contains the information about the nonpoint interaction due to the structures of hadrons. The value of $\Lambda$ is near 1 GeV which is the typical scale of nonperturbative QCD interaction. In this work, we shall treat the cutoff $\Lambda$ in the form factors as a parameter varying in a much wider range 0.8-4.8 GeV to see if the BS equation has solutions. We also vary the parameters $\zeta_1$ and $\mu_1$ in the reasonable range, in order to check if the results are sensitive to the effective coupling constants.

In our calculation, we choose to work in the rest frame of the system in which $P$ = ($M$,0). We take the averaged masses of the mesons from the PDG \cite{Tanabashi:2018oca}, $m_{D}$ = 1868.04 MeV, $m_{D_1}$ = 2422.00 MeV, $m_\rho$ = 775.49 MeV, $m_\omega$ = 782.65 MeV, $m_\sigma$ = 600 MeV, and $m_N=938.27$ MeV. Then, we can explore whether there are bound states with $I(J^P)=0(1^-)$ and $I(J^P)=1(1^-)$ for the $D_1D$ system by solving the BS equation.

\subsection{The exchanged mesons without considering $\sigma$}
\label{NR-off-sigma}
In this subsection, we will present the numerical results for the ground $D_1D$ molecular state when only $\rho$ and $\omega$ are exchanged between the constituents. From our calculations, only $Y(4260)$ could be a $D_1D$ molecular state, while $Z_2^+(4250)$ cannot. $Y(4260)$ could be an $I=0$ $D_1D$ molecular state with $\Lambda_M$, $\Lambda_D$ and $\Lambda_E$ in the range (1373, 1321) MeV, (1946, 1865) MeV and (1392, 1329) MeV for different parameters $\zeta_1$ and $\mu_1$, respectively.

\begin{figure}[htbp]
\centering
\subfigure[]{
\begin{minipage}[t]{0.3\linewidth}
\centering
\includegraphics[width=1.9in]{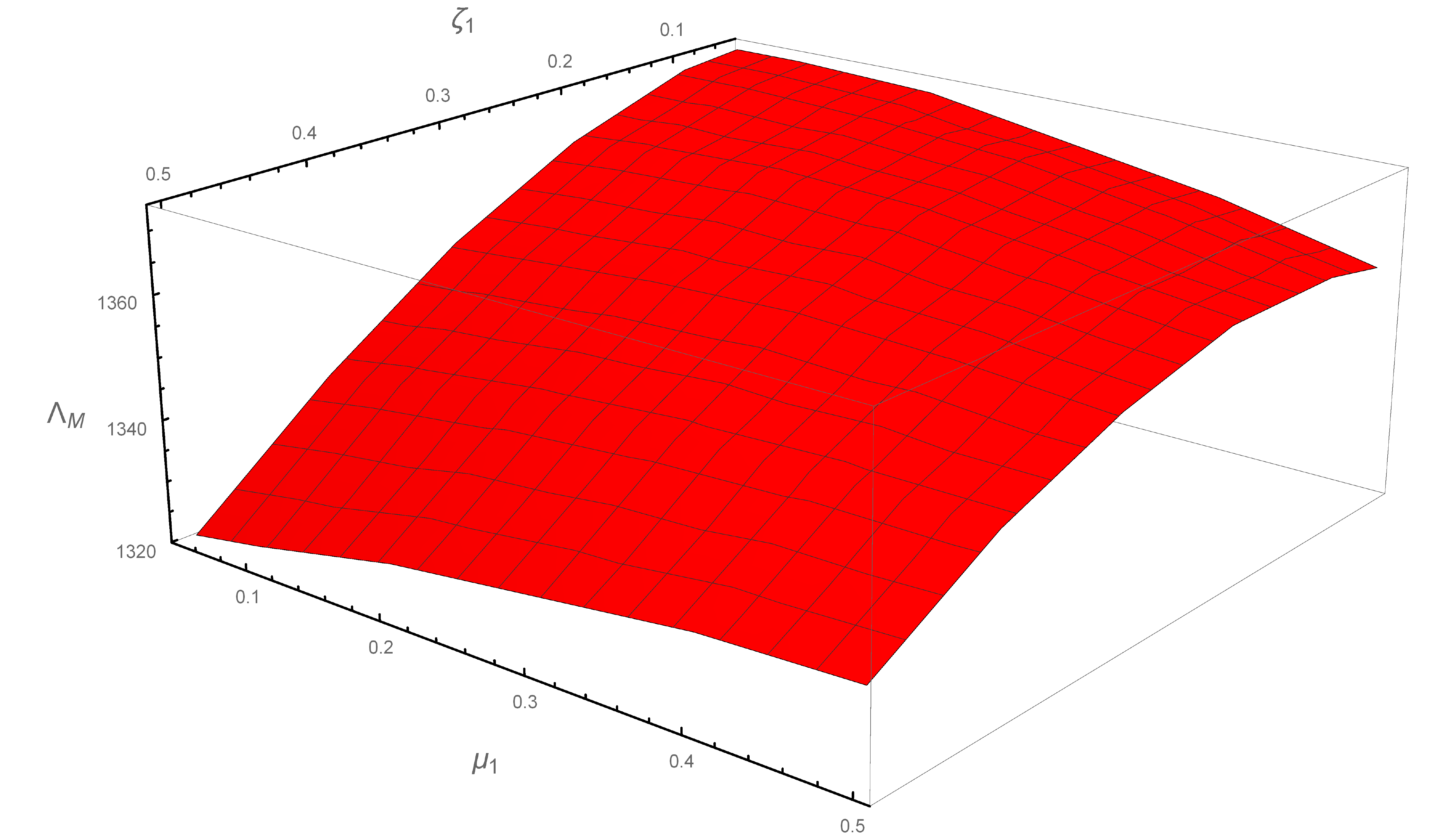}
\end{minipage}%
}%
\subfigure[]{
\begin{minipage}[t]{0.3\linewidth}
\centering
\includegraphics[width=1.9in]{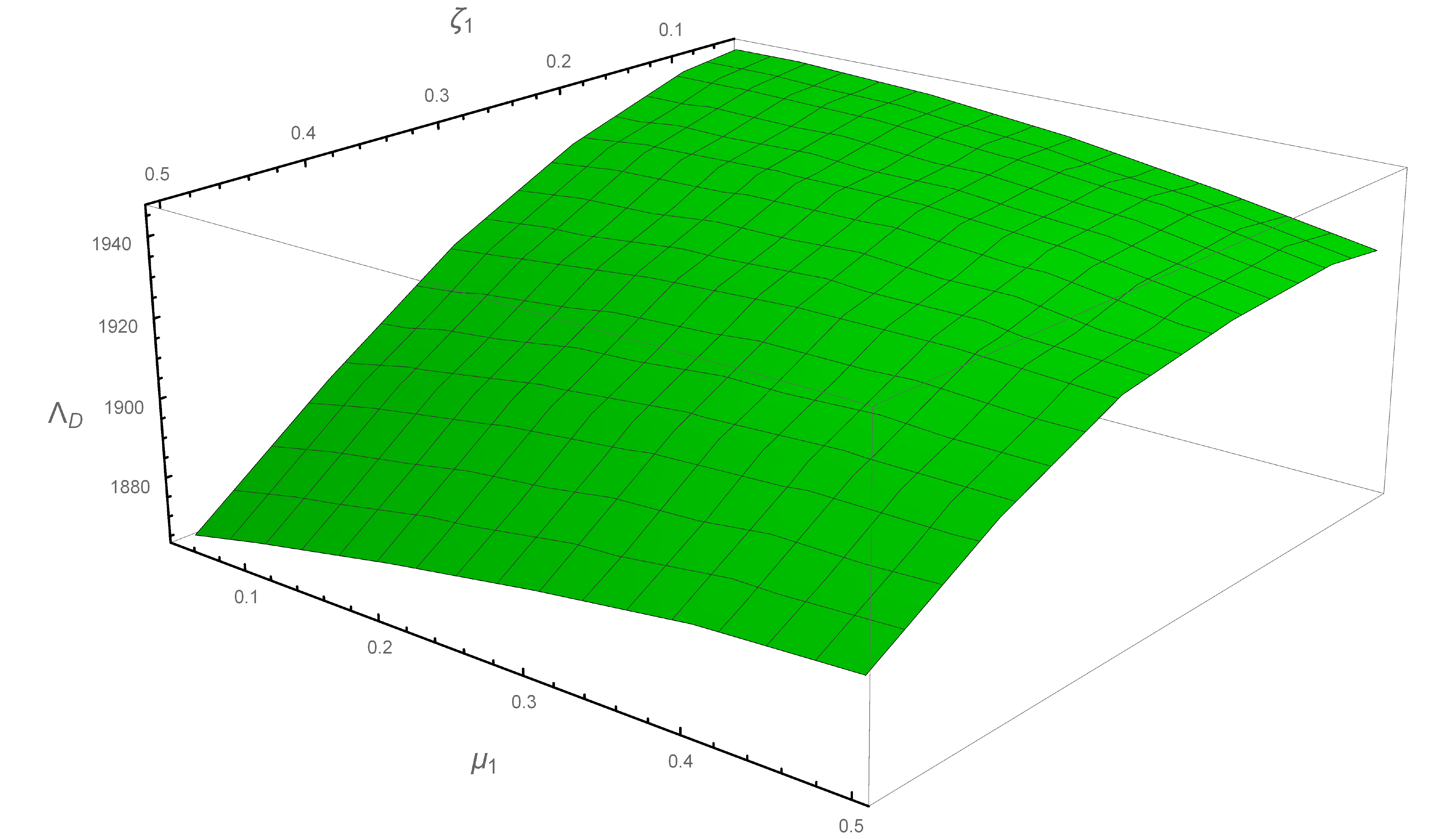}
\end{minipage}
}%
\subfigure[]{
\begin{minipage}[t]{0.3\linewidth}
\centering
\includegraphics[width=1.9in]{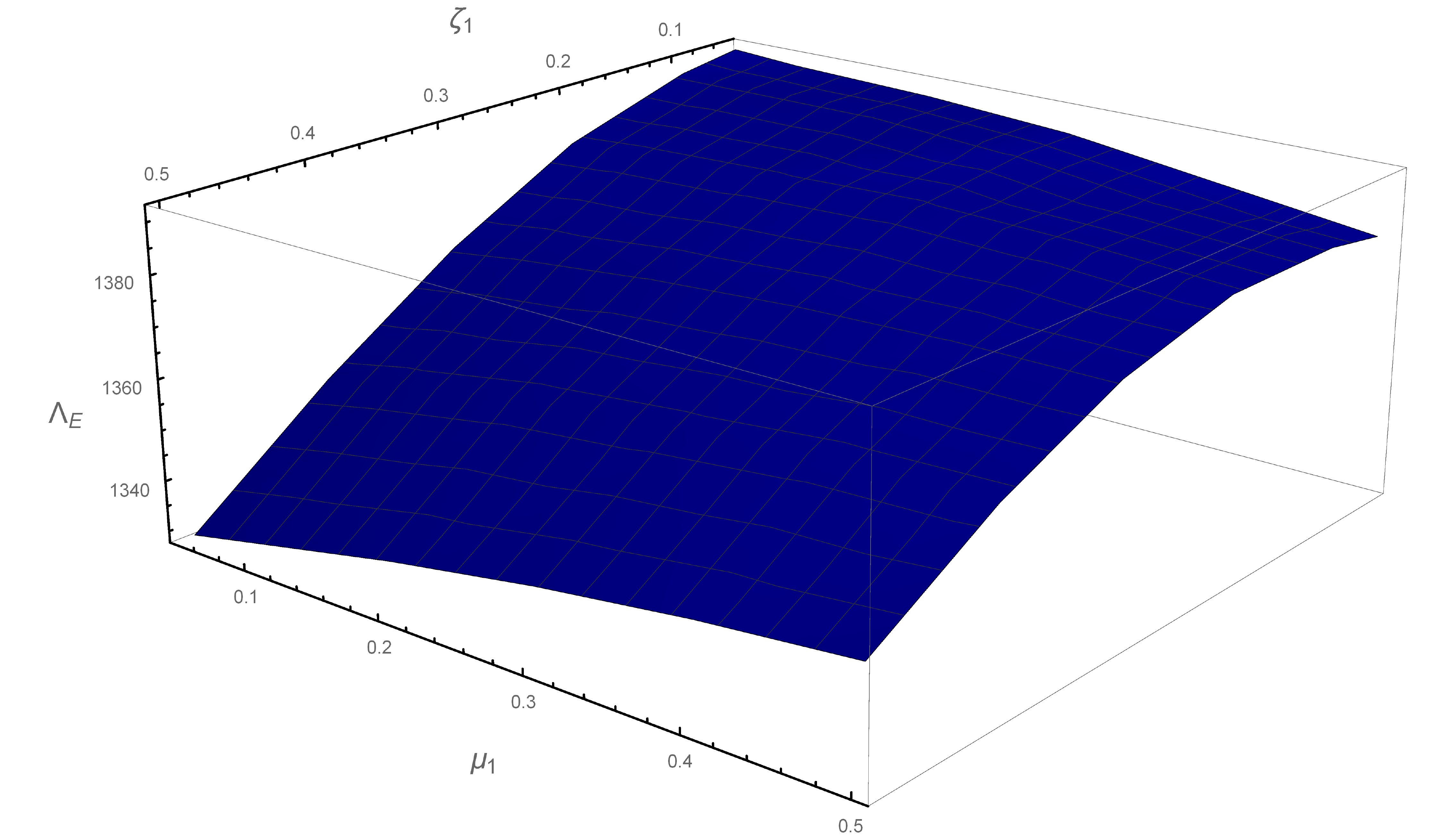}
\end{minipage}%
}%
\centering
\caption{Dependence of the cutoff $\Lambda_M$ (a), $\Lambda_D$ (b) and $\Lambda_E$ (b) of $Y(4260)$ without considering $\sigma$ meson exchanged between the constituents on parameters $\zeta_1$ and $\mu_1$.}
\label{4260Lambda-off-sigma}
\end{figure}

The 3D cutoff $\Lambda_M$, $\Lambda_D$ and $\Lambda_E$ graphics are presented in Figure \ref{4260Lambda-off-sigma}. From these figures, we see that the cutoff show small variations with respect to the changes in the parameters $\zeta_1$ and $\mu_1$ and the cutoff $\Lambda$ is more sensitive to $\zeta_1$ than $\mu_1$. We can also intuitively find the cutoff $\Lambda$ in the dipole form factor are larger than those in the monopole and exponential form factors while they vary in reasonable regions. Therefore, $Y(4260)$ could be a molecular state in appropriate effective coupling constants and the cutoff.

\begin{figure}[htbp]
\centering
\subfigure[]{
\begin{minipage}[t]{0.3\linewidth}
\centering
\includegraphics[width=1.7in]{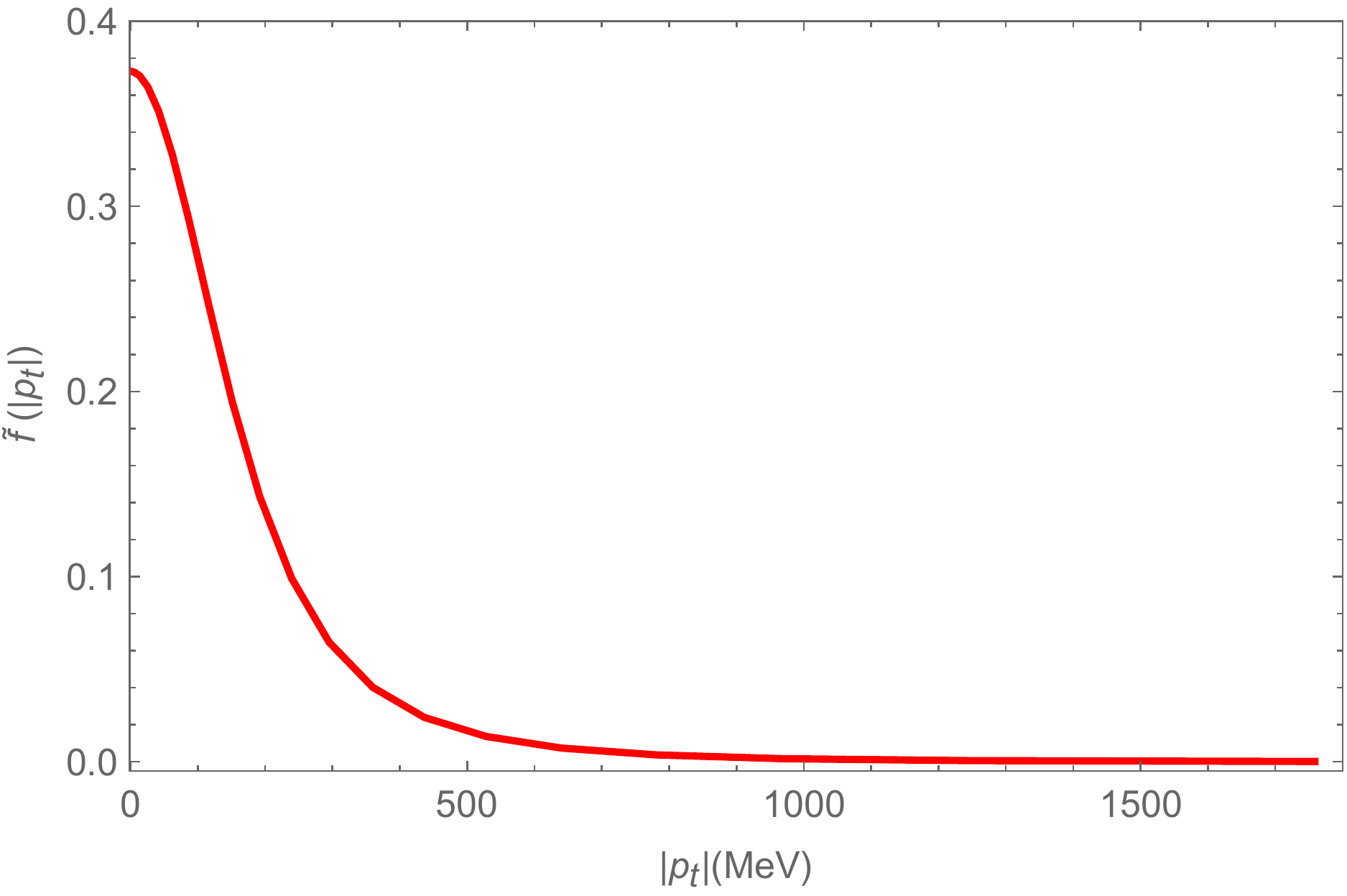}
\end{minipage}%
}%
\subfigure[]{
\begin{minipage}[t]{0.3\linewidth}
\centering
\includegraphics[width=1.7in]{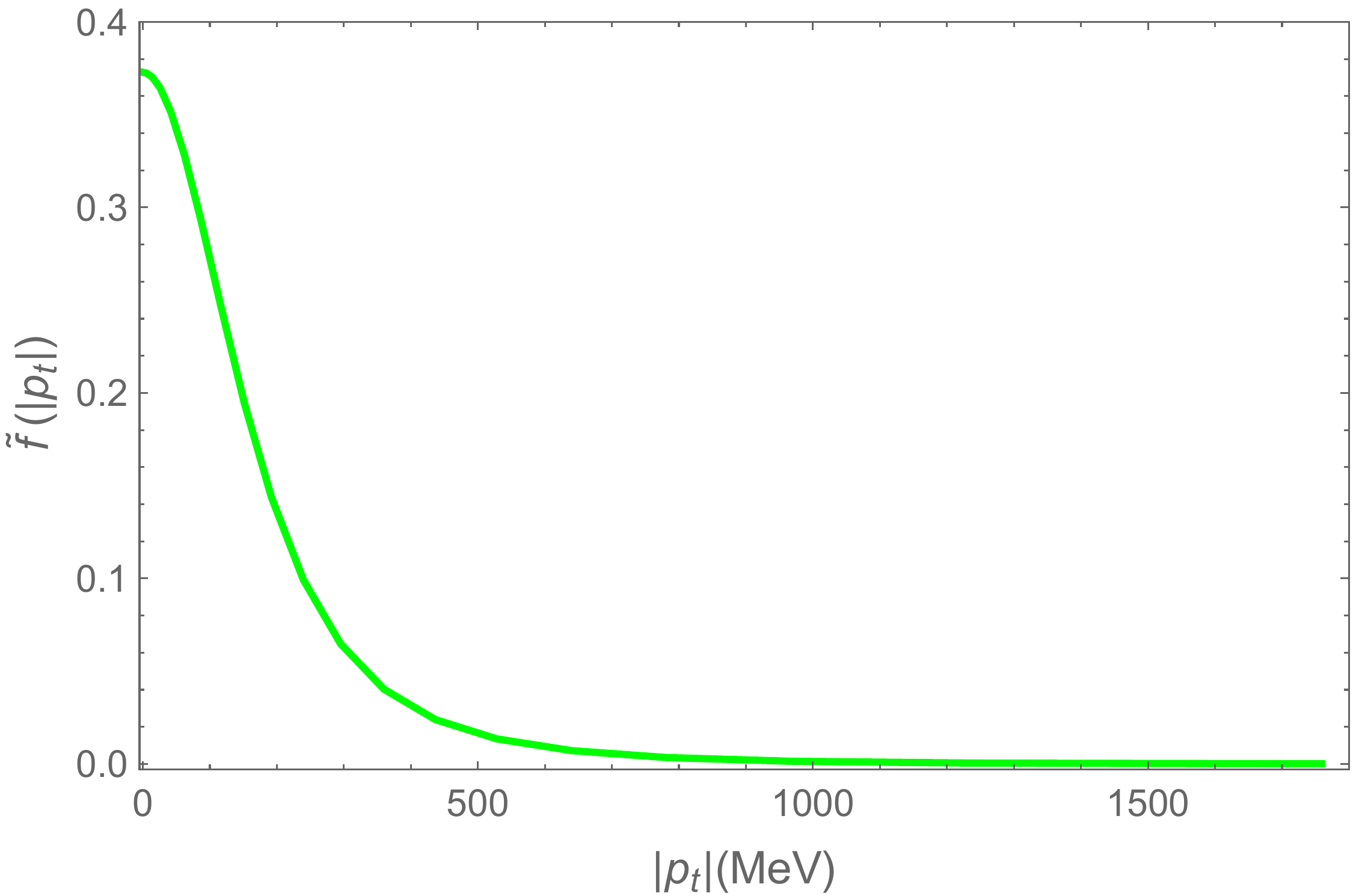}
\end{minipage}
}%
\subfigure[]{
\begin{minipage}[t]{0.3\linewidth}
\centering
\includegraphics[width=1.7in]{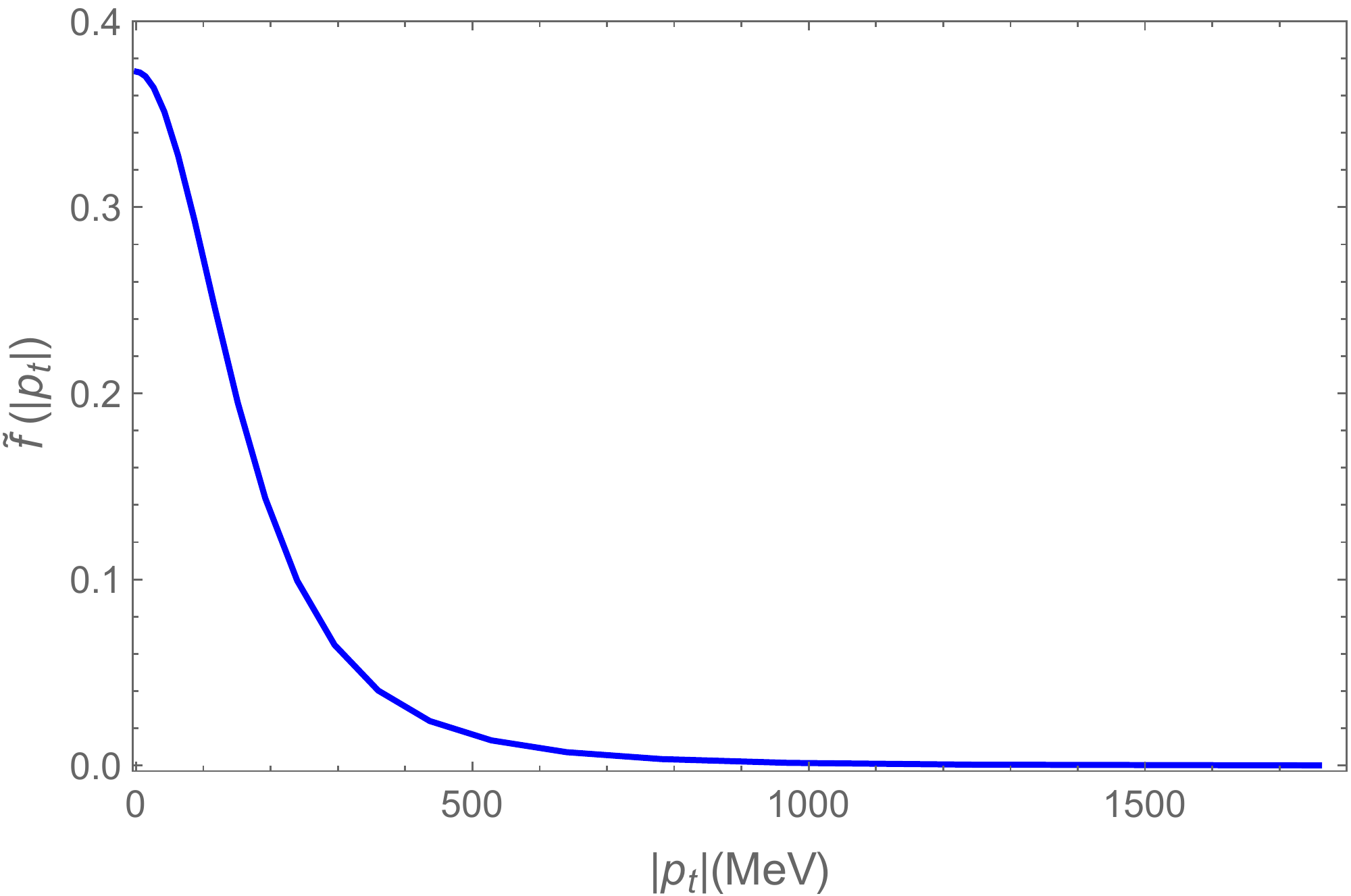}
\end{minipage}%
}%
\centering
\caption{Numerical results of the wave function $\tilde{f}(|p_t|)$ for $Y(4260)$ without considering $\sigma$ meson contribution in the $D_1\bar{D}$ molecular picture with (a) the monopole form factor, (b) the dipole form factor, and (c) the exponential form factor for the parameters $\mu=0.1$ $\mathrm{GeV}^{-1}$ and $\zeta=0.1$.}
\label{4260-bound-state}
\end{figure}

In Figure \ref{4260-bound-state}, we present the numerical results of the wave functions for three different form factors with parameters $\mu=0.1$ $\mathrm{GeV}^{-1}$ and $\zeta=0.1$. From these figures, we see that the numerical results of the wave functions with monopole, dipole and exponential form factors are almost the same. The numerical results of the wave functions corresponding to other parameter values show the same situation. This indicates that the different forms of the form factor have little effect on the wave function.

In Refs. \cite{Cleven:2013mka,Qin:2016spb}, the authors interpreted $Y(4260)$ as a $D_1D$ bound state, but they also predicted a significantly smaller mass of about 4.22 GeV. Soon after $Y(4220)$ was observed by BES$\mathrm{\uppercase\expandafter{\romannumeral3}}$ Collaboration \cite{Ablikim:2016qzw}. The mass of $Y(4220)$ is about 58 MeV below the threshold for $D_1D$. Therefore, we vary the binding energy $E_b$ in the region from -60 to 0 MeV trying to find all the possible solutions.

From our calculations, the $D_1D$ system can not be an $I=1$ bound state. Hence like $Z_2^+(4250)$ can not be an $I=1$ $D_1D$ molecule. The reason for that is the exchanges of $\rho$ and $\omega$ cancel each other for the quantum number $I=1$.

\begin{table}[h]
\begin{spacing}{0.9}
\centering
\caption{
The numerical results for the possible $I=0$ $D_1D$ molecular state without considering $\sigma$ meson contribution for the monopole, dipole, and exponential form factors.}
\begin{tabular*}{\textwidth}{@{\extracolsep{\fill}}|c|cccccc|}
\hline
  $E_b$  ($\mathrm{MeV}$)     &  -60       &  -50        & -40         & -30         &  -20        &  -10  \\
\hline
$\Lambda_M$ ($\mathrm{MeV}$)  &  1531-1455 &  1483-1414  &  1431-1370  &  1373-1321  &  1305-1262  &  1217-1184   \\
$\Lambda_D$ ($\mathrm{MeV}$)  &  2199-2084 &  2123-2019  &  2040-1946  &  1946-1864  &  1834-1766  &  1686-1635     \\
$\Lambda_E$ ($\mathrm{MeV}$)  &  1594-1507 &  1534-1454  &  1467-1396  &  1391-1329  &  1300-1248  &  1180-1139  \\
\hline
\end{tabular*}\label{ND1Dpossible-results}
\end{spacing}
\end{table}

\begin{figure}[htbp]
\centering
\subfigure[]{
\begin{minipage}[t]{0.3\linewidth}
\centering
\includegraphics[width=1.9in]{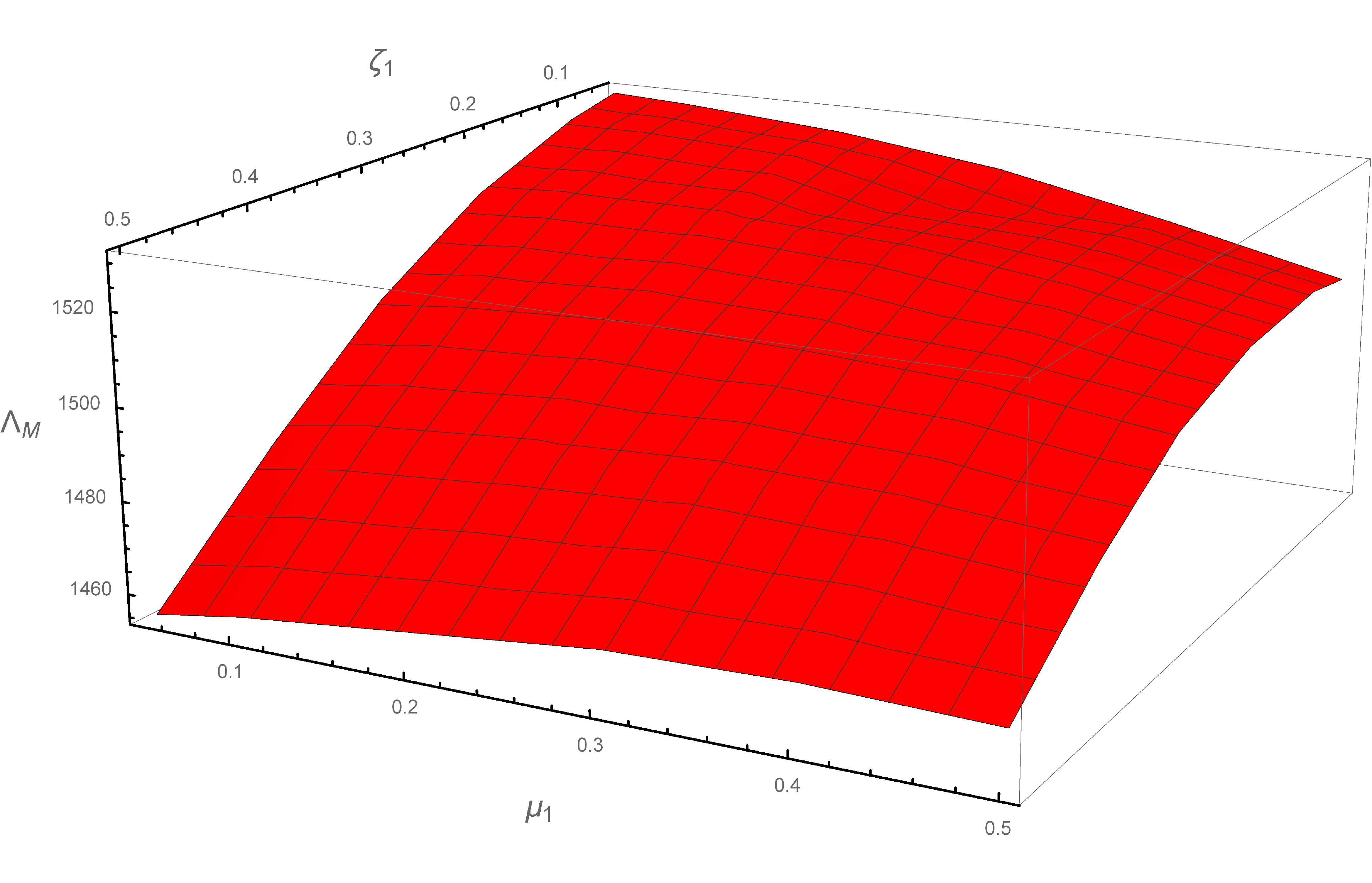}
\end{minipage}%
}%
\subfigure[]{
\begin{minipage}[t]{0.3\linewidth}
\centering
\includegraphics[width=1.9in]{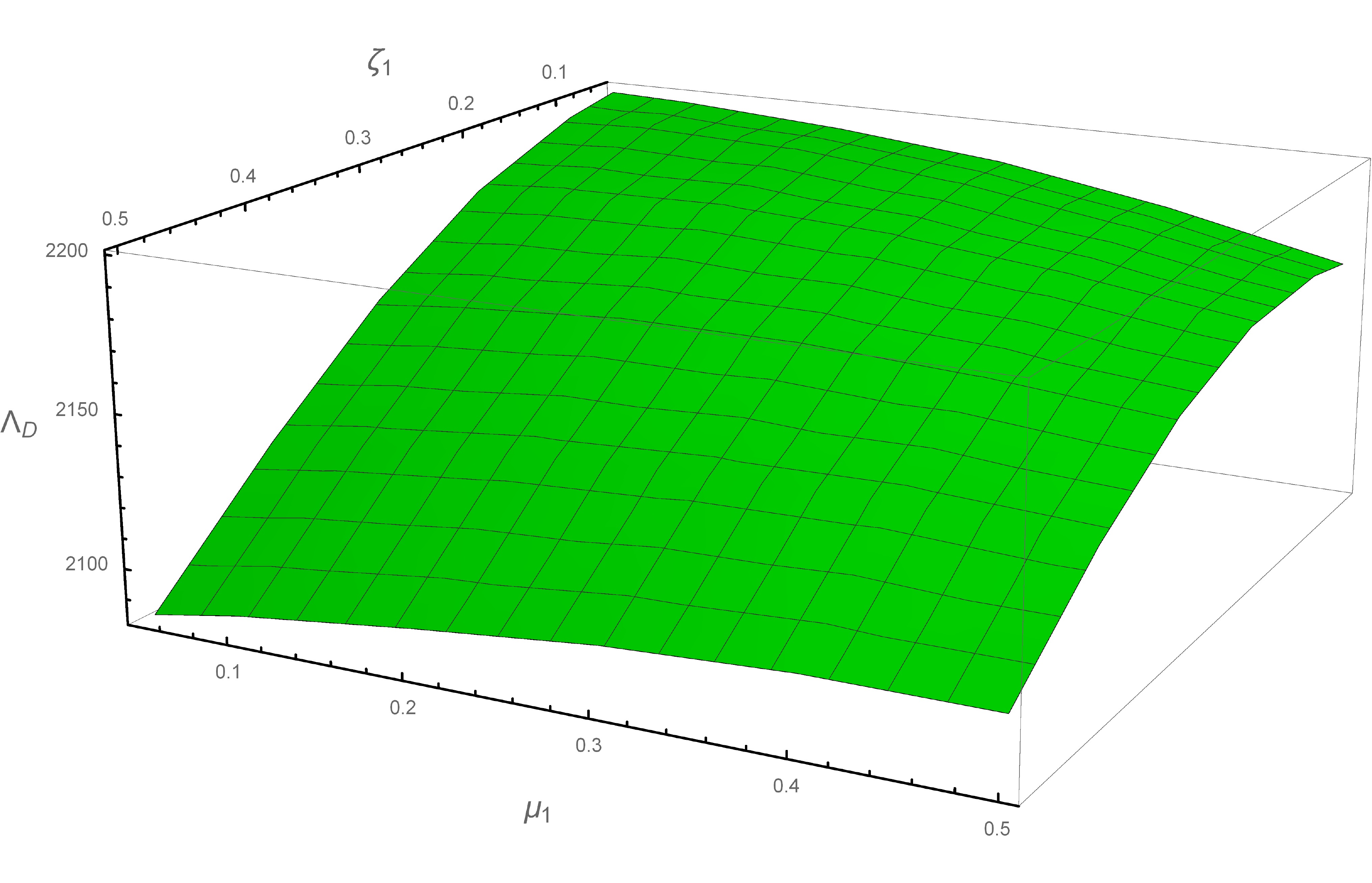}
\end{minipage}
}%
\subfigure[]{
\begin{minipage}[t]{0.3\linewidth}
\centering
\includegraphics[width=1.9in]{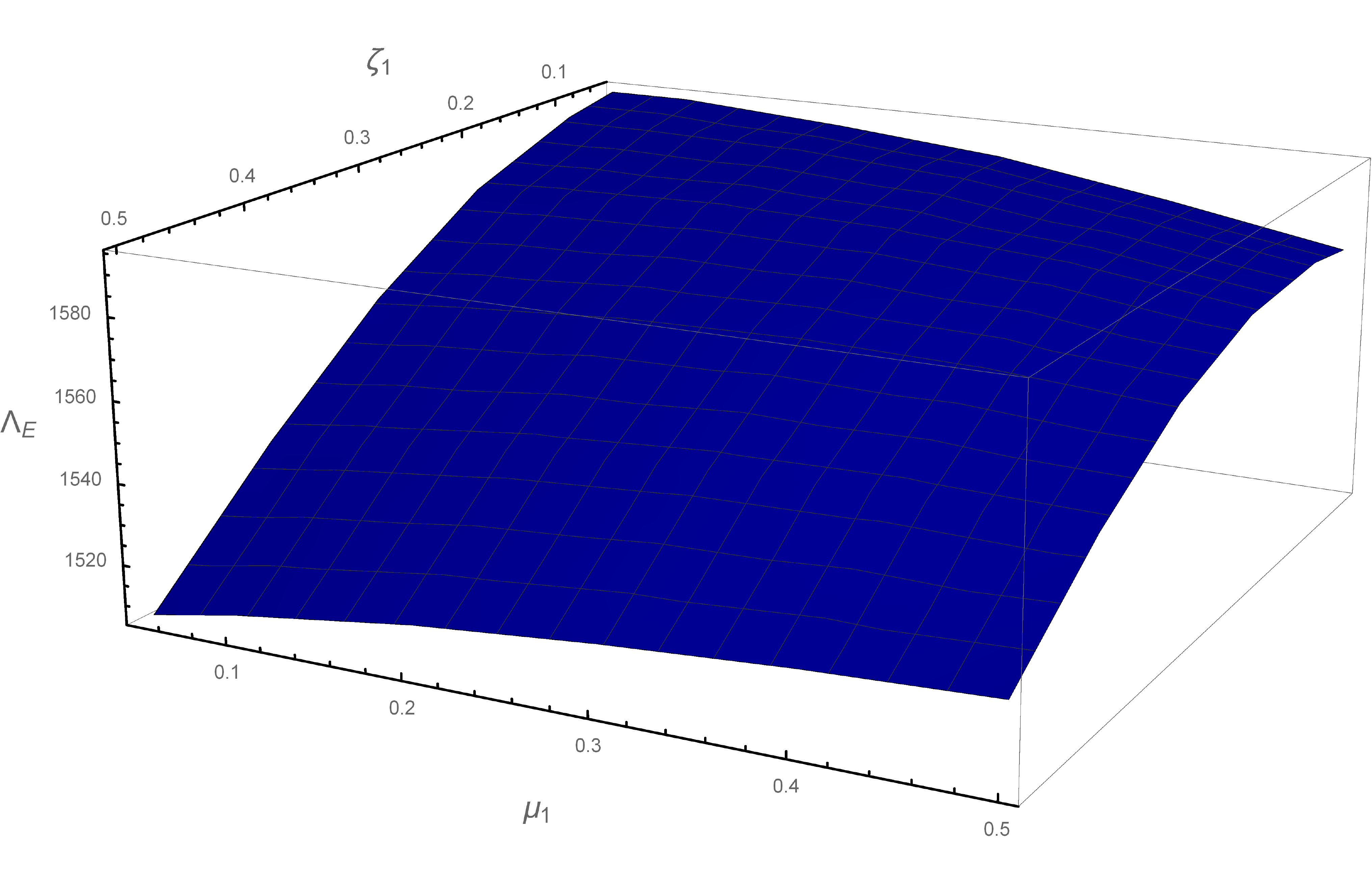}
\end{minipage}%
}%
\centering
\caption{Dependence of the cutoff $\Lambda_M$(a), $\Lambda_D$(b) and $\Lambda_E$(b) for the $I=0$ $D_1D$ bound state without considering $\sigma$ meson contribution on the parameters $\zeta_1$ and $\mu_1$ with $E_b=-60$ MeV.}
\label{ND1DLabda}
\end{figure}

For the $I=0$ $D_1D$ system, we find several regions for the cutoffs. The results are listed in Table \ref{ND1Dpossible-results}, from which we can see the cutoffs $\Lambda$ in three difference form factors are in reasonable ranges and all of them become smaller with the increasing of the binding energy. We depict the 3D graphics (see Figure \ref{ND1DLabda}) showing the variation of the cutoffs with respect to the parameters $\zeta_1$ and $\mu_1$ when $E_b=-60$ MeV. For other 3D graphics with different $E_b$, the variation trend is the same as that in Figure \ref{ND1DLabda}. We can conclude that the $D_1D$ system could be a molecular state. Our result is consistent with the meson exchange model based on the heavy meson chiral perturbation theory \cite{Ding:2008gr} and the chiral quark model in which the isoscalar channel is found to be easier to bind than the isovector channel for the same components \cite{Swanson:2006st}.

\begin{figure}[htbp]
\centering
\subfigure[]{
\begin{minipage}[t]{0.3\linewidth}
\centering
\includegraphics[width=1.9in]{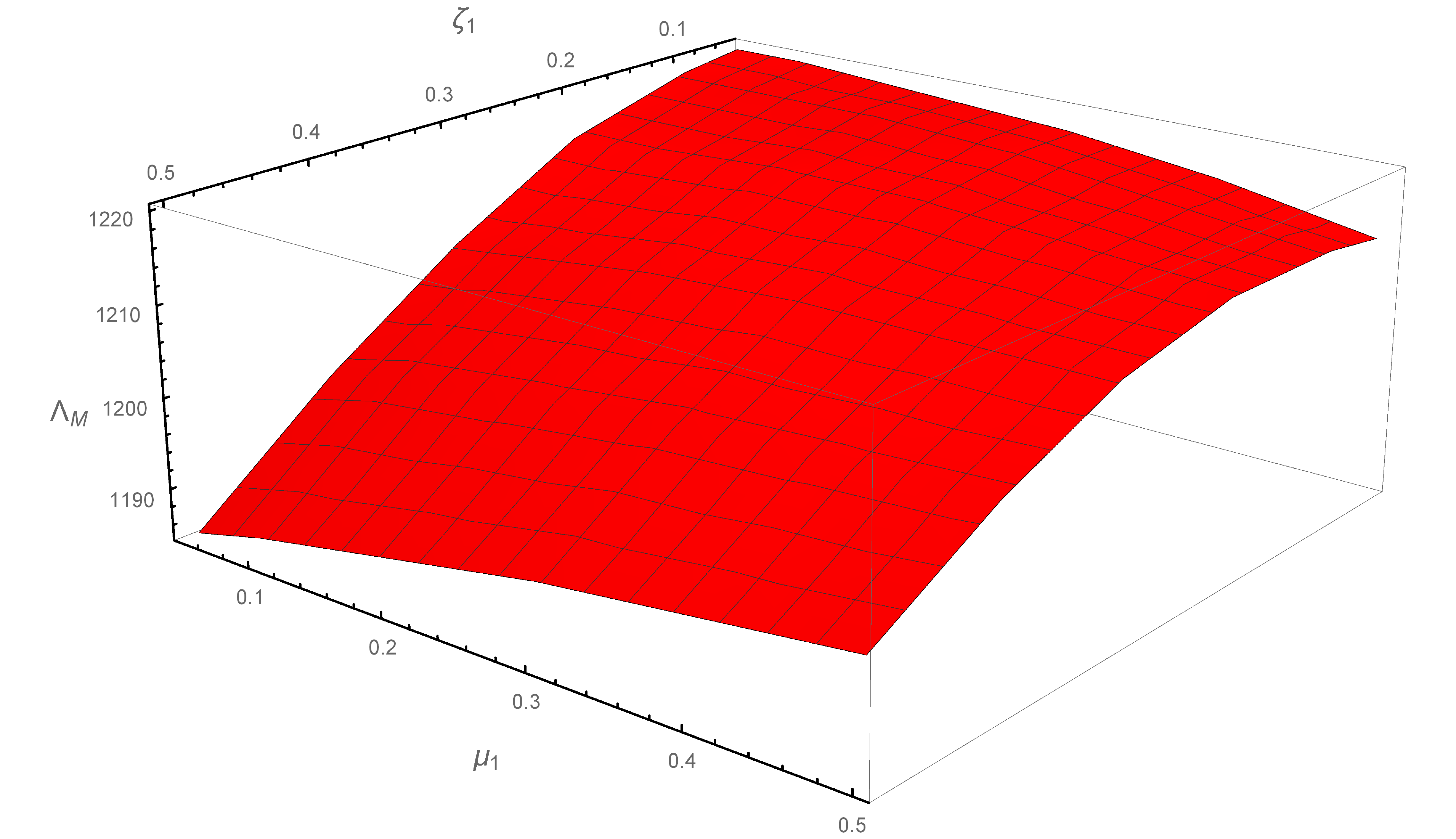}
\end{minipage}%
}%
\subfigure[]{
\begin{minipage}[t]{0.3\linewidth}
\centering
\includegraphics[width=1.9in]{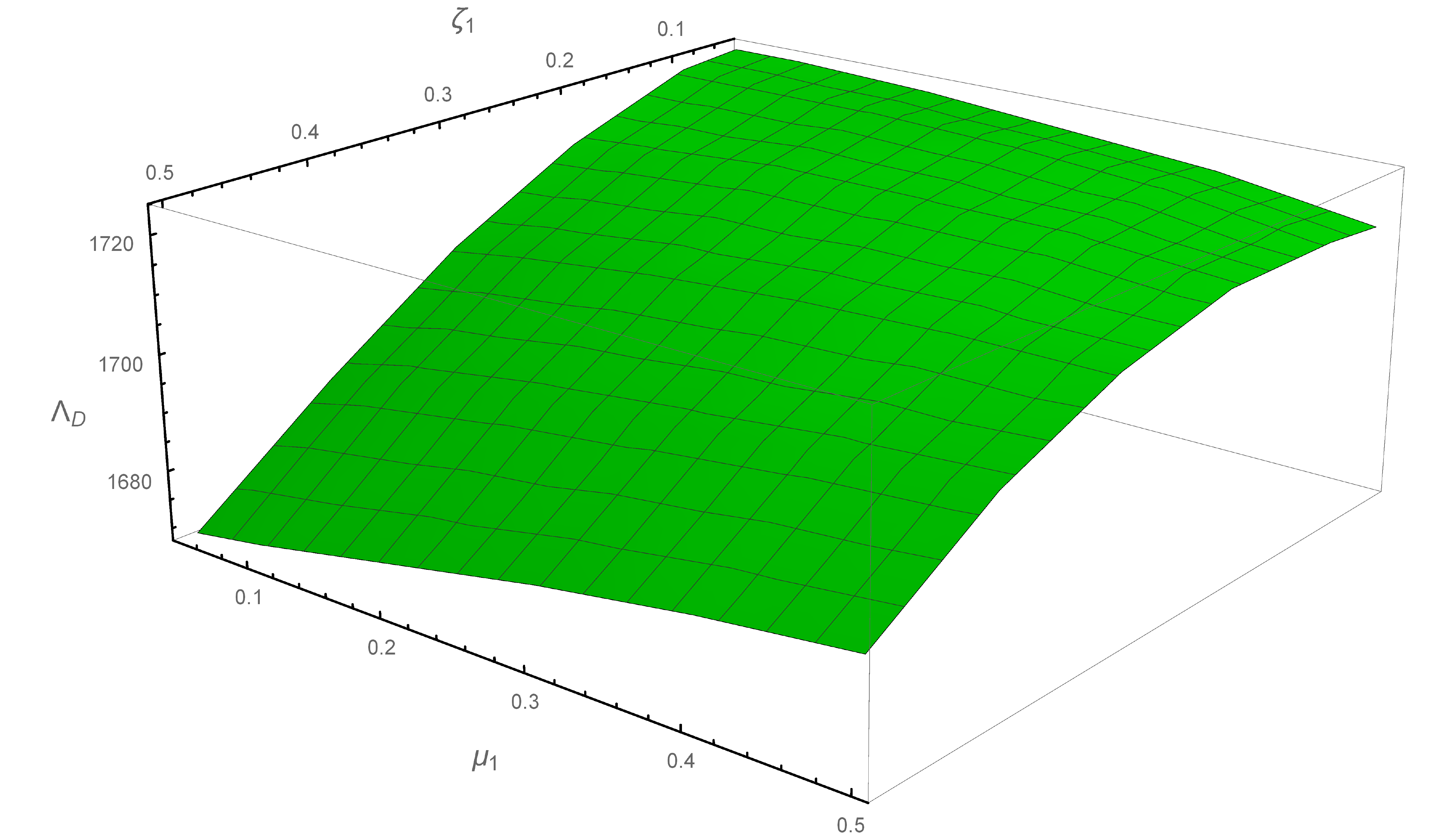}
\end{minipage}
}%
\subfigure[]{
\begin{minipage}[t]{0.3\linewidth}
\centering
\includegraphics[width=1.9in]{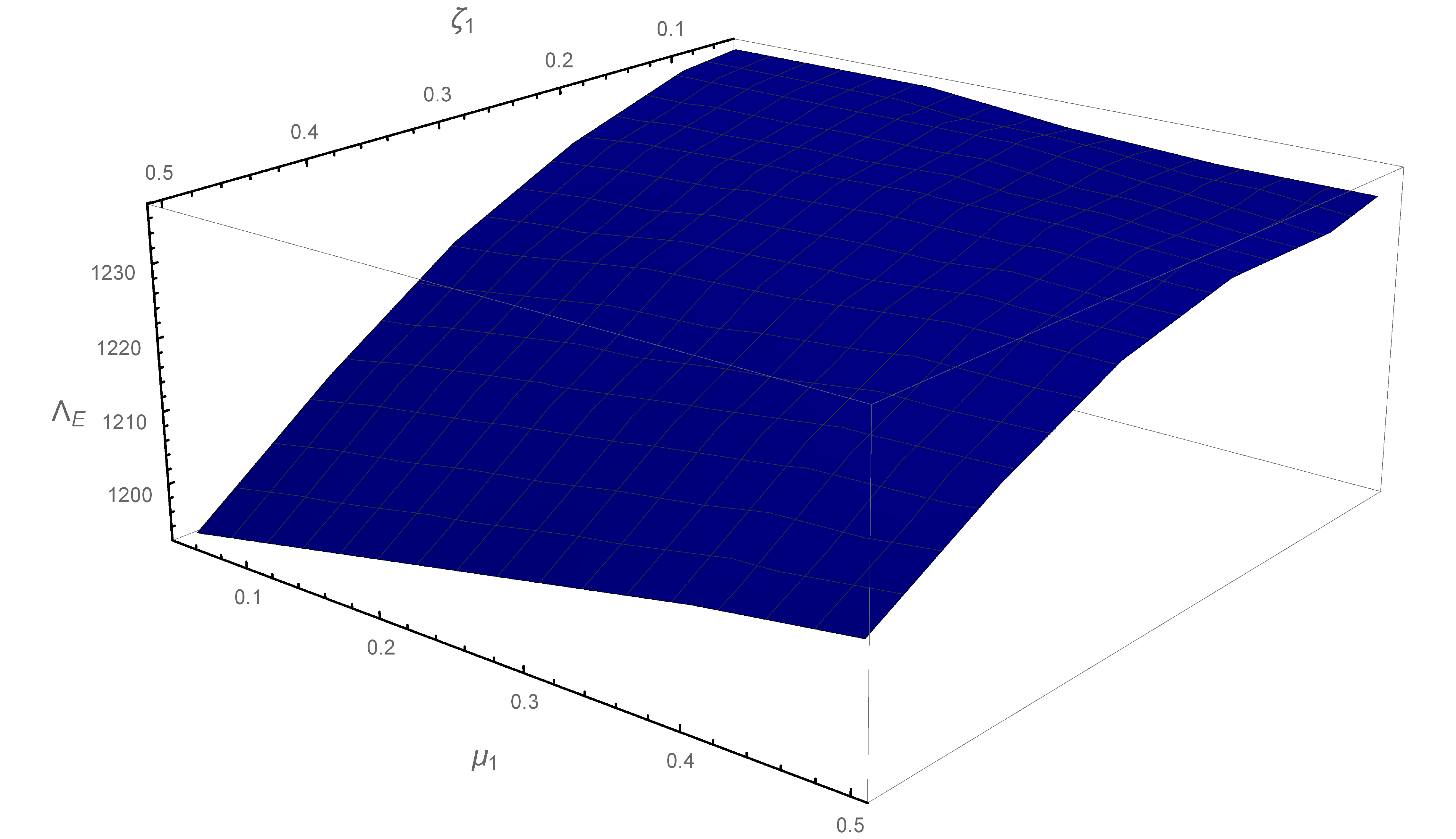}
\end{minipage}%
}%
\centering
\caption{Dependence of the cutoff $\Lambda_M$(a), $\Lambda_D$(b) and $\Lambda_E$(b) for the $I=0$ $B_1B$ bound state without considering $\sigma$ meson contribution on the parameters $\zeta_1$ and $\mu_1$ with $E_b=-60$ MeV.}
\label{B1BLabda}
\end{figure}

The same procedure can be easily extended to study the bottom analogs of $Y(4260)$ and $Z_2^+(4250)$, by simply replacing the charm quark and antiquark with the bottom quark and antiquark, respectively. The masses of the bottom mesons are $m_{B_1}=5726.0$ MeV and $m_{B}=5279.4$ MeV \cite{Tanabashi:2018oca}. Due to the heavier mass of the bottom meson the kinematic term is relatively small, resulting in the bottom system to form the molecular state more easier. We use the same set of parameters as in the $D_1D$ system. With these parameters the $I=0$ $B_1B$ bound state always exists with the reasonable cutoff. The numerical results for the $B_1B$ system are shown in Table \ref{NB1Bpossible-results}. Similar to the $D_1D$ system, all the cutoffs $\Lambda_M$, $\Lambda_D$ and $\Lambda_E$ become smaller with the increasing of the binding energy. The variations of the cutoffs with respect to the parameters $\zeta_1$ and $\mu_1$ when $E_b=-60$ MeV are presented in Figure \ref{B1BLabda}.

\begin{table}[h]
\begin{spacing}{0.9}
\centering
\caption{
The numerical results for the possible $I=0$ $B_1B$ molecular state without considering $\sigma$ meson contribution for the monopole, dipole, and exponential form factors.}
\begin{tabular*}{\textwidth}{@{\extracolsep{\fill}}|c|cccccc|}
\hline
  $E_b$     ($\mathrm{MeV}$) &  -60       &  -50        & -40         & -30         &  -20        &  -10  \\
\hline
$\Lambda_M$ ($\mathrm{MeV}$) &  1220-1185 &  1185-1155  &  1148-1121  &  1107-1085  &  1060-1043  &  1002-990   \\
$\Lambda_D$ ($\mathrm{MeV}$) &  1724-1668 &  1664-1614  &  1559-1556  &  1526-1490  &  1441-1413  &  1333-1313     \\
$\Lambda_E$ ($\mathrm{MeV}$) &  1237-1193 &  1187-1148  &  1132-1098  &  1071-1042  &   998-975   &  903-887  \\
\hline
\end{tabular*}\label{NB1Bpossible-results}
\end{spacing}
\end{table}

\subsection{The exchanged meosons with considering $\sigma$}

\begin{figure}[htbp]
\centering
\subfigure[]{
\begin{minipage}[t]{0.3\linewidth}
\centering
\includegraphics[width=1.7in]{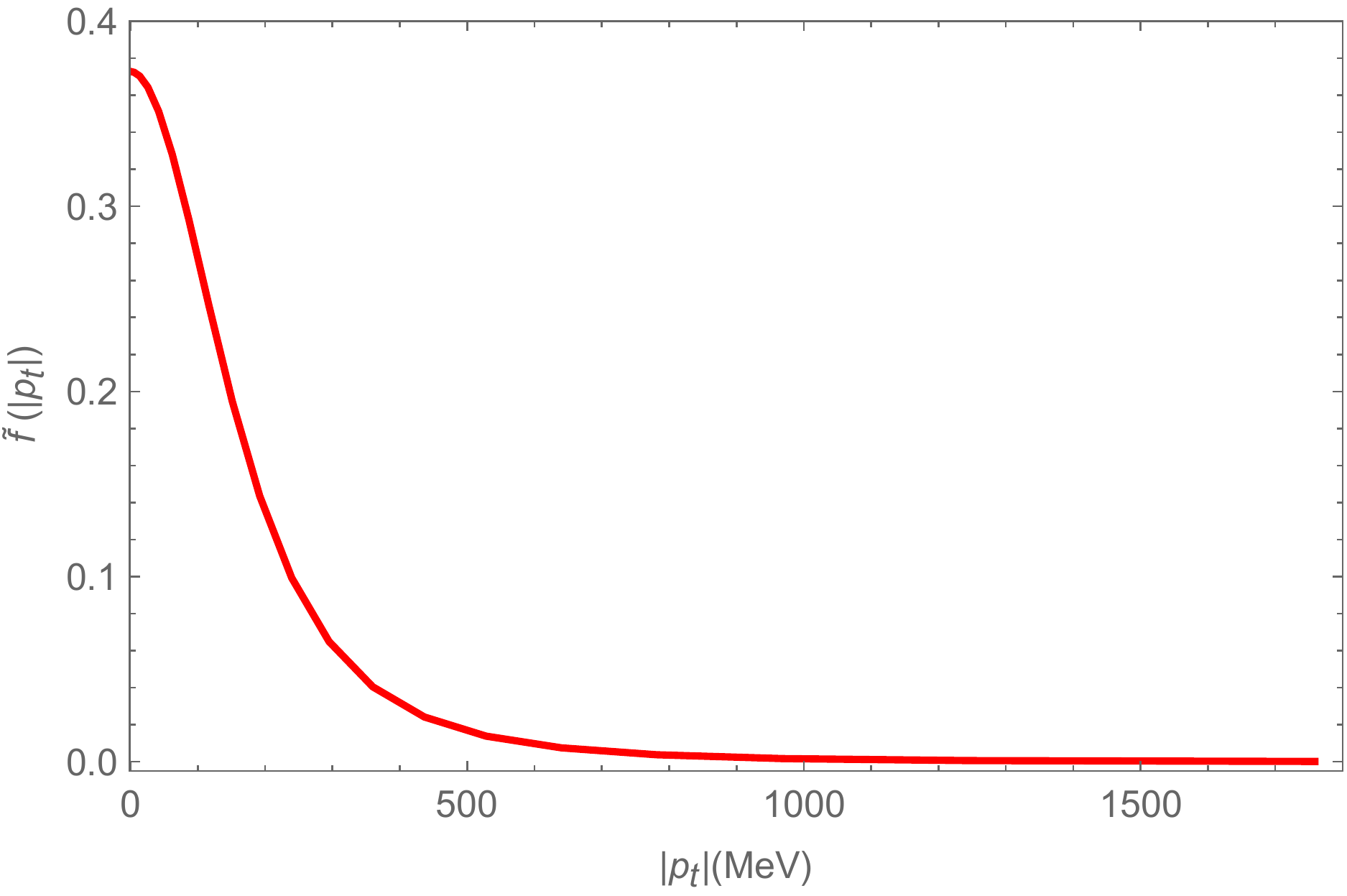}
\end{minipage}%
}%
\subfigure[]{
\begin{minipage}[t]{0.3\linewidth}
\centering
\includegraphics[width=1.7in]{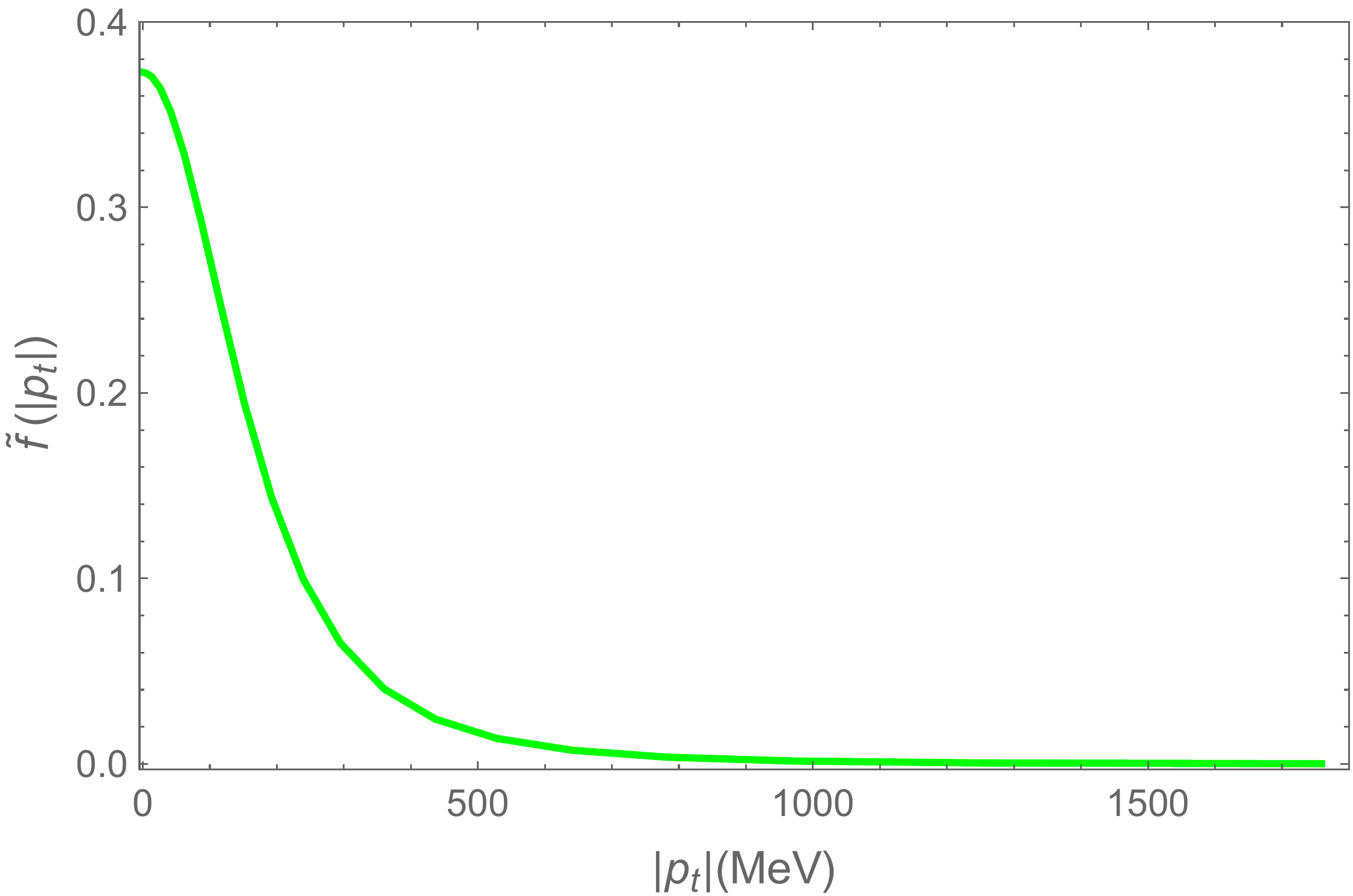}
\end{minipage}
}%
\subfigure[]{
\begin{minipage}[t]{0.3\linewidth}
\centering
\includegraphics[width=1.7in]{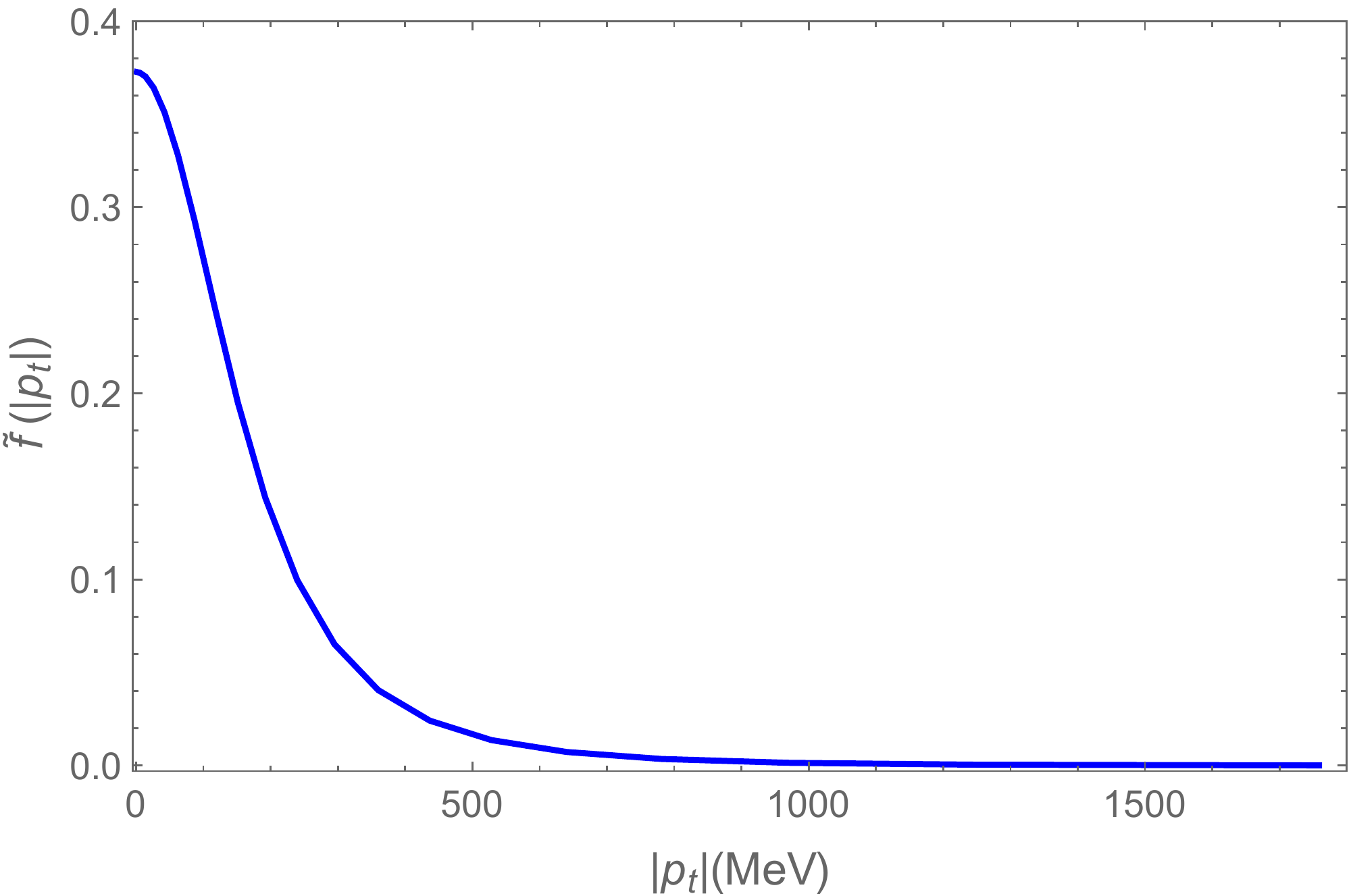}
\end{minipage}%
}%
\centering
\caption{Numerical results of the wave function $\tilde{f}(|p_t|)$ for $Y(4260)$ in the $D_1\bar{D}$ molecular picture with (a) the monopole form factor, (b) the dipole form factor, and (c) the exponential form factor for the parameters $\mu=0.1$ $\mathrm{GeV}^{-1}$ and $\zeta=0.1$.}
\label{W-bound-state}
\end{figure}

In this subsection, we will consider all the contributions of $\sigma$, $\rho$ and $\omega$ mesons exchange in the kernel. From our studied, the contribution from $\sigma$ meson exchange has little effect on the results. $Y(4260)$ could be a $I=0$  $D_1D$ molecular state with $\Lambda_M$, $\Lambda_D$ and $\Lambda_E$ in the range (1362, 1312) MeV, (1930, 1852) MeV and (1380, 1320) MeV for different parameters $\zeta_1$ and $\mu_1$, respectively, while $Z_2^+(4250)$ also cannot. The variation trend of cutoffs $\Lambda$ is same as in Subsec. \ref{NR-off-sigma}. In Figure \ref{W-bound-state}, we present the numerical results of the wave functions for three different form factors with parameters $\mu=0.1$ $\mathrm{GeV}^{-1}$ and $\zeta=0.1$.

\begin{table}[h]
\begin{spacing}{0.9}
\centering
\caption{
The numerical results for the possible $D_1D$ molecular state with $I=0$ for the monopole, dipole, and exponential form factors.}
\begin{tabular*}{\textwidth}{@{\extracolsep{\fill}}|c|cccccc|}
\hline
  $E_b$  ($\mathrm{MeV}$)     &  -60       &  -50        & -40         & -30         &  -20        &  -10  \\
\hline
$\Lambda_M$ ($\mathrm{MeV}$)  &  1514-1443 &  1468-1403  &  1418-1360  &  1362-1312  &  1295-1254  &  1209-1178   \\
$\Lambda_D$ ($\mathrm{MeV}$)  &  2176-2067 &  2103-2003  &  2022-1932  &  1930-1852  &  1820-1755  &  1676-1626     \\
$\Lambda_E$ ($\mathrm{MeV}$)  &  1578-1495 &  1519-1443  &  1454-1386  &  1380-1320  &  1291-1240  &  1172-1133  \\
\hline
\end{tabular*}\label{D1Dpossible-results}
\end{spacing}
\end{table}

In Table \ref{D1Dpossible-results}, we give the numerical results of the monopole, dipole, and exponential form factors for the possible $I=0$ $D_1D$ molecular state with the binding energy in the region (-60, 0) MeV. In the region where the binding energy is between -60 and 0 MeV, we still have not found the existence of the $D_1D$ molecular state with $I=0$. The numerical results for the $I=0$ $B_1B$ system are shown in Table \ref{B1Bpossible-results}.

\begin{table}[h]
\begin{spacing}{0.9}
\centering
\caption{
The numerical results for the possible $I=0$ $B_1B$ molecular state for the monopole, dipole, and exponential form factors.}
\begin{tabular*}{\textwidth}{@{\extracolsep{\fill}}|c|cccccc|}
\hline
  $E_b$     ($\mathrm{MeV}$) &  -60       &  -50        & -40         & -30         &  -20        &  -10  \\
\hline
$\Lambda_M$ ($\mathrm{MeV}$) &  1211-1178 &  1177-1148  &  1141-1116  &  1100-1080  &  1055-1038  &  997-986   \\
$\Lambda_D$ ($\mathrm{MeV}$) &  1712-1658 &  1653-1606  &  1589-1548  &  1518-1483  &  1434-1407  &  1328-1308     \\
$\Lambda_E$ ($\mathrm{MeV}$) &  1229-1186 &  1180-1142  &  1126-1093  &  1065-1037  &   993-971   &  900-884  \\
\hline
\end{tabular*}\label{B1Bpossible-results}
\end{spacing}
\end{table}

Compared the related results in this Subsection with Subsec. \ref{NR-off-sigma}, we can find the values of cutoffs $\Lambda$ becomes a little smaller when considering the $\sigma$ exchange contribution. The contribution of $\sigma$ exchange for the wave function can be negligible.  As studied in Ref. \cite{Ke:2012gm}, when we enhanced the coupling constant $g_{\sigma}$ by several times, there existence $I=1$ $D_1D$ molecular states. Therefore, the small coupling constant $g_\sigma$ determines that including the $\sigma$ contribution cannot make a substantial change to the kernel.

\section{summary}
\label{su}
In this work, we studied whether $Y(4260)$ and $Z_2^+(4250)$ could be a $D_1D$ molecular state in the Bethe-Salpeter equation approach. In our model, we applied the ladder and instantaneous approximations to obtain the kernel containing one-particle-exchange diagrams and introduced three different form factors (the monopole form factor, the dipole form factor, and the exponential form factor) since the constituent particles and the exchanged particles are not pointlike. The cutoff $\Lambda$ introduced in the form factors reflects the effects of the structures of interacting particles. Since $\Lambda$ is controlled by nonperturbative QCD and cannot be determined accurately, we let it vary in a reasonable range within which we try to find possible bound states of the $D_1D$ system. Since the two effective coupling constants $\mu_1$ and $\zeta_1$ have not been determined we varied them in larger ranges, i.e. [0.05, 0.5] $\mathrm{GeV}^{-1}$ and [0.05, 0.5], respectively.

Our numerical results indicated that when the parameters are within reasonable ranges the $D_1D$ system can form an $I=0$ molecular state but cannot form an $I=1$ molecular state because the contribution of $\sigma$ meson exchange is small. In other words, $Y(4260)$ could be accommodated as a $D_1D$ molecule. However, the existence of $Z_2^+(4250)$ as a molecule requires that the coupling constant $g_\sigma$ be enhanced by several times. Consequently, the interpretation of $Z_2^+(4250)$ as a $D_1D$ molecule is disfavored. Its structure should be studied further.

The bottom analogs of $Y(4260)$ and $Z_2^+(4250)$ were also studied. Similar to the $D_1D$ system, the $B_1B$ system with $I=0$ also can form a molecular state but cannot for $I=1$. We expect forthcoming experimental measurements to test our model for the $D_1D$ and $B_1B$ systems.

\acknowledgments
This work was supported by National Natural Science Foundation of China (Projects No. 11775024, No.11575023, No.11605150 and No.11947001), the Ningbo Natural Science Foundation (No.2019A610067) and K.C.Wong Magna Fund in Ningbo University.



\begin{thebibliography}{99}

\bibitem{Aubert:2005rm}
B.~Aubert \textit{et al.} [BaBar],
Observation of a broad structure in the $\pi^+ \pi^- J/\psi$ mass spectrum around 4.26-GeV/c$^2$,
Phys.\ Rev.\ Lett.\  \textbf{95}, 142001 (2005)
[arXiv:hep-ex/0506081 [hep-ex]].

\bibitem{He:2006kg}
Q.~He \textit{et al.} [CLEO],
Confirmation of the $Y(4260)$ resonance production in ISR,
Phys.\ Rev.\ D \textbf{74}, 091104 (2006)
[arXiv:hep-ex/0611021 [hep-ex]].

\bibitem{Yuan:2007sj}
C.~Yuan \textit{et al.} [Belle],
Measurement of $e^+ e^- \rightarrow \pi^+ \pi^- J/\psi$ cross-section via initial state radiation at Belle,
Phys.\ Rev.\ Lett.\  \textbf{99}, 182004 (2007)
[arXiv:0707.2541 [hep-ex]].

\bibitem{Ablikim:2013mio}
M.~Ablikim \textit{et al.} [BESIII],
Observation of a Charged Charmoniumlike Structure in $e^+e^-\rightarrow \pi^+\pi^- J/\psi$ at $\sqrt{s}$ =4.26  GeV,
Phys.\ Rev.\ Lett.\  \textbf{110}, 252001 (2013)
[arXiv:1303.5949 [hep-ex]].

\bibitem{Tanabashi:2018oca}
M.~Tanabashi \textit{et al.} [Particle Data Group],
Review of Particle Physics,
Phys.\ Rev.\ D \textbf{98} no.3, 030001 (2018)

\bibitem{Ablikim:2016qzw}
M.~Ablikim \textit{et al.} [BESIII],
Precise measurement of the $e^+e^-\to \pi^+\pi^-J/\psi$ cross section at center-of-mass energies from 3.77 to 4.60 GeV,
Phys.\ Rev.\ Lett.\  \textbf{118} no.9, 092001 (2017)
[arXiv:1611.01317 [hep-ex]].

\bibitem{Ablikim:2018vxx}
M.~Ablikim \textit{et al.} [BESIII],
Evidence of a resonant structure in the $e^+e^-\to \pi^+D^0D^{*-}$ cross section between 4.05 and 4.60 GeV,
Phys.\ Rev.\ Lett.\  \textbf{122} no.10, 102002 (2019)
[arXiv:1808.02847 [hep-ex]].

\bibitem{Abe:2006fj}
K.~Abe \textit{et al.} [Belle],
Measurement of the near-threshold $e^+ e^- \rightarrow D^{(*)\pm} D^{(*)\mp}$ cross section using initial-state radiation,
Phys.\ Rev.\ Lett.\  \textbf{98}, 092001 (2007)
[arXiv:hep-ex/0608018 [hep-ex]].

\bibitem{Pakhlova:2008zza}
G.~Pakhlova \textit{et al.} [Belle],
Measurement of the near-threshold $e^+ e^- \rightarrow D \bar{D}$ cross section using initial-state radiation,
Phys.\ Rev.\ D \textbf{77}, 011103 (2008)
[arXiv:0708.0082 [hep-ex]].

\bibitem{Aubert:2009aq}
B.~Aubert \textit{et al.} [BaBar],
Exclusive Initial-State-Radiation Production of the $D \bar{D}$, $D^\ast \bar{D}^\ast$, and $D* \bar{D}^\ast$ Systems,
Phys.\ Rev.\ D \textbf{79}, 092001 (2009)
[arXiv:0903.1597 [hep-ex]].

\bibitem{CroninHennessy:2008yi}
D.~Cronin-Hennessy \textit{et al.} [CLEO],
Measurement of Charm Production Cross Sections in $e^+e^-$ Annihilation at Energies between 3.97 and 4.26-GeV,
Phys.\ Rev.\ D \textbf{80}, 072001 (2009)
[arXiv:0801.3418 [hep-ex]].


\bibitem{Brambilla:2010cs}
N.~Brambilla \textit{et al.},
Heavy Quarkonium: Progress, Puzzles, and Opportunities,
Eur.\ Phys.\ J.\ C \textbf{71}, 1534 (2011)
[arXiv:1010.5827 [hep-ph]].


        %
        %

\bibitem{Zhu:2005hp}
  S.~L.~Zhu,
The Possible interpretations of $Y(4260)$,
Phys.\ Lett.\ B \textbf{625}, 212 (2005)
[arXiv:hep-ph/0507025 [hep-ph]].

\bibitem{Kou:2005gt}
E.~Kou and O.~Pene,
Suppressed decay into open charm for the $Y(4260)$ being an hybrid,
Phys.\ Lett.\ B \textbf{631}, 164-169 (2005)
[arXiv:hep-ph/0507119 [hep-ph]].



          %
          %
\bibitem{Ebert:2008kb}
D.~Ebert, R.~Faustov and V.~Galkin,
Excited heavy tetraquarks with hidden charm,
Eur.\ Phys.\ J.\ C \textbf{58}, 399-405 (2008)
[arXiv:0808.3912 [hep-ph]].

\bibitem{Ali:2011qi}
A.~Ali and W.~Wang,
Production of the Exotic $1^{--}$ Hadrons $\phi(2170)$, X(4260) and $Y_b(10890)$ at the LHC and Tevatron via the Drell-Yan Mechanism,
Phys.\ Rev.\ Lett.\  \textbf{106}, 192001 (2011)
[arXiv:1103.4587 [hep-ph]].

\bibitem{Dias:2012ek}
J.~Dias, R.~Albuquerque, M.~Nielsen and C.~Zanetti,
$Y(4260)$ as a mixed charmonium-tetraquark state,
Phys. Rev. D \textbf{86}, 116012 (2012)
[arXiv:1209.6592 [hep-ph]].


\bibitem{Albuquerque:2015nwa}
R.~Albuquerque, M.~Nielsen and C.~Zanetti,
Production of the $Y(4260)$ state in B meson decay,
Phys. Lett. B \textbf{747}, 83-87 (2015)
[arXiv:1502.00119 [hep-ph]].

\bibitem{Wang:2016mmg}
Z.~G.~Wang,
Tetraquark state candidates: $Y(4260)$, $Y(4360)$, $Y(4660)$ and $Z_c(4020/4025)$,
Eur. Phys. J. C \textbf{76} no.7, 387 (2016)
[arXiv:1601.05541 [hep-ph]].



\bibitem{Liu:2005ay}
X.~Liu, X.~Q.~Zeng and X.~Q.~Li,
Possible molecular structure of the newly observed $Y(4260)$,
Phys. Rev. D \textbf{72}, 054023 (2005)
[arXiv:hep-ph/0507177 [hep-ph]].

\bibitem{Yuan:2005dr}
C.~Z.~Yuan, P.~Wang and X.~H.~Mo,
Phys. Lett. B \textbf{634}, 399-402 (2006)
[arXiv:hep-ph/0511107 [hep-ph]].

\bibitem{Albuquerque:2008up}
R.~Albuquerque and M.~Nielsen,
QCD sum rules study of the $J^{PC} = 1^{--}$ charmonium Y mesons,
Nucl. Phys. A \textbf{815}, 53-66 (2009)
[arXiv:0804.4817 [hep-ph]].

\bibitem{Albuquerque:2011ix}
R.~M.~Albuquerque, M.~Nielsen and R.~Rodrigues da Silva,
Exotic $1^{--}$ States in QCD Sum Rules,
Phys. Rev. D \textbf{84}, 116004 (2011)
[arXiv:1110.2113 [hep-ph]].


\bibitem{Ding:2008gr}
G.~J.~Ding,
Are $Y(4260)$ and $Z^+_2$ $D_1 D$ or $D_0 D^\ast$ Hadronic Molecules?,
Phys. Rev. D \textbf{79}, 014001 (2009)
[arXiv:0809.4818 [hep-ph]].

\bibitem{Liu:2013vfa}
  X.~H.~Liu and G.~Li,
Exploring the threshold behavior and implications on the nature of $Y(4260)$ and $Z_c(3900)$,
Phys. Rev. D \textbf{88}, 014013 (2013)
[arXiv:1306.1384 [hep-ph]].

\bibitem{Cleven:2013mka}
  M.~Cleven, Q.~Wang, F.~K.~Guo, C.~Hanhart, U.~G.~Mei\ss{}ner and Q.~Zhao,
$Y(4260)$ as the first $S$-wave open charm vector molecular state?,
Phys. Rev. D \textbf{90} no.7, 074039 (2014)
[arXiv:1310.2190 [hep-ph]].

\bibitem{Qin:2016spb}
  W.~Qin, S.~R.~Xue and Q.~Zhao,
  Production of $Y(4260)$ as a hadronic molecule state of $\bar{D}D_1 +c.c.$ in $e^+e^-$ annihilations,
Phys. Rev. D \textbf{94} no.5, 054035 (2016)
[arXiv:1605.02407 [hep-ph]].

\bibitem{Cleven:2016qbn}
  M.~Cleven and Q.~Zhao,
 Cross section line shape of $e^+e^-\to\chi_{c0}\omega$ around the $Y(4260)$ mass region,
Phys. Lett. B \textbf{768}, 52-56 (2017)
[arXiv:1611.04408 [hep-ph]].

\bibitem{Xue:2017xpu}
  S.~R.~Xue, H.~J.~Jing, F.~K.~Guo and Q.~Zhao,
  Disentangling the role of the $Y(4260)$ in $e^+e^-\to D^*\bar{D}^*$ and $D_s^*\bar{D}_s^*$ via line shape studies,
Phys. Lett. B \textbf{779}, 402-408 (2018)
[arXiv:1708.06961 [hep-ph]].

\bibitem{Chen:2019mgp}
  Y.~H.~Chen, L.~Y.~Dai, F.~K.~Guo and B.~Kubis,
  Nature of the $Y(4260)$: A light-quark perspective,
  Phys. Rev. D \textbf{99} no.7, 074016 (2019)
[arXiv:1902.10957 [hep-ph]].




\bibitem{MartinezTorres:2009xb}
  A.~Martinez Torres, K.~P.~Khemchandani, D.~Gamermann and E.~Oset,
  The $Y(4260)$ as a $J/\psi K \bar{K}$ system,
  Phys. Rev. D \textbf{80}, 094012 (2009)
[arXiv:0906.5333 [nucl-th]].




\bibitem{Qiao:2005av}
  C.~F.~Qiao,
  One explanation for the exotic state $Y(4260)$,
Phys. Lett. B \textbf{639}, 263-265 (2006)
[arXiv:hep-ph/0510228 [hep-ph]].




\bibitem{Mizuk:2008me}
  R.~Mizuk {\it et al.} [Belle Collaboration],
  Observation of two resonance-like structures in the $\pi^+ \chi_{c1}$ mass distribution in exclusive $\bar{B}_0 \rightarrow K^- \pi^+ \chi_{c1}$ decays,
Phys. Rev. D \textbf{78}, 072004 (2008)
[arXiv:0806.4098 [hep-ex]].

\bibitem{Lees:2011ik}
  J.~P.~Lees {\it et al.} [BaBar Collaboration],
  Search for the $Z_1(4050)^+$ and $Z_2(4250)^+$ states in $\bar B^0 \to \chi_{c1} K^- \pi^+$ and $B^+ \to \chi_{c1} K^0_S \pi^+$,
  Phys. Rev. D \textbf{85}, 052003 (2012)
[arXiv:1111.5919 [hep-ex]].

\bibitem{Sbordone:2013exb}
F.~Sbordone,
Study of the decay $B^0 \to \chi_{c1} K^+ \pi^-$ and search of exotic resonances at LHCb,
CERN-THESIS-2013-294.

\bibitem{Lee:2008tz}
  S.~H.~Lee, K.~Morita and M.~Nielsen,
  Width of exotics from QCD sum rules: Tetraquarks or molecules?,
  Phys. Rev. D \textbf{78}, 076001 (2008)
[arXiv:0808.3168 [hep-ph]].


\bibitem{Wang:2008af}
  Z.~G.~Wang,
  Another tetraquark structure in the $\pi^+ \chi_{c1}$ invariant mass distribution,''
  Eur. Phys. J. C \textbf{62}, 375-382 (2009)
[arXiv:0807.4592 [hep-ph]].

\bibitem{Deng:2015lca}
  C.~Deng, J.~Ping, H.~Huang and F.~Wang,
  Systematic study of Z$_c^+$ family from a multiquark color flux-tube model,
  Phys. Rev. D \textbf{92} no.3, 034027 (2015)
[arXiv:1507.06408 [hep-ph]].

\bibitem{Nakamura:2019emd}
  S.~X.~Nakamura,
  ``Triangle singularities in $\bar{B}^0\to \chi_{c1}K^-\pi^+$ relevant to $Z_1(4050)$ and $Z_2(4250)$,''
  Phys.\ Rev.\ D {\bf 100}, no. 1, 011504 (2019)




\bibitem{Lurie}
David Lurie, Particles and Fields (Interscience Publishers, New York, 1968), Chap. 9.

\bibitem{Bardeen:2003kt}
  W.~A.~Bardeen, E.~J.~Eichten and C.~T.~Hill,
  Phys. Rev. D \textbf{68}, 054024 (2003)
[arXiv:hep-ph/0305049 [hep-ph]].

\bibitem{Liu:2008xz}
  X.~Liu, Y.~R.~Liu, W.~Z.~Deng and S.~L.~Zhu,
  $Z^+(4430)$ as a $D'_1 D^\ast(D_1 D^\ast)$  molecular state,''
  Phys. Rev. D \textbf{77}, 094015 (2008)
doi:10.1103/PhysRevD.77.094015
[arXiv:0803.1295 [hep-ph]].

\bibitem{Wang:2019aoc}
F.~Wang, R.~Chen, Z.~Liu and X.~Liu,
Possible triple-charm molecular pentaquarks from $\Xi_{cc}D_1/\Xi_{cc}D_2^*$ interactions,
Phys. Rev. D \textbf{99} no.5, 054021 (2019)
[arXiv:1901.01542 [hep-ph]].

\bibitem{Casalbuoni:1996pg}
  R.~Casalbuoni, A.~Deandrea, N.~Di Bartolomeo, R.~Gatto, F.~Feruglio and G.~Nardulli,
  Phenomenology of heavy meson chiral Lagrangians,
Phys. Rept. \textbf{281}, 145-238 (1997)
[arXiv:hep-ph/9605342 [hep-ph]].

\bibitem{Gross:1982nz}
  F.~Gross,
  The Relativistic Few Body Problem. 1. Two-Body Equations,
Phys. Rev. C \textbf{26}, 2203-2225 (1982)

\bibitem{Theussl:1999xq}
  L.~Theussl and B.~Desplanques,
  Crossed boson exchange contribution and Bethe-Salpeter equation,
 Few Body Syst. \textbf{30}, 5-19 (2001)
[arXiv:nucl-th/9908007 [nucl-th]].


\bibitem{Guo:2007mm}
  X.~H.~Guo and X.~H.~Wu,
  Studying the scalar bound states of $K \bar{K}$ system in Bethe-Salpeter formalism,
 Phys. Rev. D \textbf{76}, 056004 (2007)
[arXiv:0704.3105 [hep-ph]].

\bibitem{Chen:2017vai}
  R.~Chen, A.~Hosaka and X.~Liu,
  Heavy molecules and one-$\sigma/\omega$-exchange model,
  Phys.\ Rev.\ D {\bf 96},  116012 (2017)
  [arXiv:1707.08306 [hep-ph]].







\bibitem{He:2014nya}
  J.~He,
  Study of the $B\bar{B}^*/D\bar{D}^*$ bound states in a Bethe-Salpeter approach,
Phys. Rev. D \textbf{90} no.7, 076008 (2014)
[arXiv:1409.8506 [hep-ph]].

\bibitem{Feng:2012zzf}
  G.~Q.~Feng and X.~H.~Guo,
  $DK$ molecule in the Bethe-Salpeter equation approach in the heavy quark limit,
  Phys. Rev. D \textbf{86}, 036004 (2012)

\bibitem{Chen:2017xat}
  R.~Chen, A.~Hosaka and X.~Liu,
  Searching for possible $\Omega_c$-like molecular states from meson-baryon interaction,
  Phys. Rev. D \textbf{97} no.3, 036016 (2018)
[arXiv:1711.07650 [hep-ph]].

\bibitem{Guo:1996jj}
  X.~H.~Guo and T.~Muta,
  Isgur-Wise function for $\Lambda_b \rightarrow \Lambda_c$ in B-S approach,
  Phys. Rev. D \textbf{54}, 4629-4634 (1996)
[arXiv:hep-ph/9706394 [hep-ph]].

\bibitem{Wang:2019ehs}
Z.~Y.~Wang, J.~J.~Qi, Q.~X.~Yu and X.~H.~Guo,
$B_{s1}(5778)$ as a $B^*\bar{K}$ molecule in the Bethe-Salpeter equation approach,
Phys. Rev. D \textbf{100}, 096009 (2019)
[arXiv:1906.09002 [hep-ph]].


\bibitem{Ke:2012gm}
H.~Ke, X.~Li, Y.~Shi, G.~Wang and X.~Yuan,
Is $Z_b(10610)$ a Molecular State?,
JHEP \textbf{04}, 056 (2012)
[arXiv:1202.2178 [hep-ph]].

\bibitem{Swanson:2006st}
  E.~S.~Swanson,
  The New heavy mesons: A Status report,
  Phys. Rept. \textbf{429}, 243-305 (2006)
[arXiv:hep-ph/0601110 [hep-ph]].





\end{thebibliography}
\end{document}